\documentclass[11pt]{article}
\usepackage{graphicx, amssymb}
\newcommand{\SetFigFont}[3]{}

\title{Decay of Solutions of the Wave Equation in the\\ Kerr Geometry}
\author{F.\ Finster\thanks{Research supported in part by the Deutsche
Forschungsgemeinschaft.}, N.\ Kamran\thanks{Research supported by NSERC grant
\# RGPIN 105490-2004.},
J.\ Smoller\thanks{Research supported in part by the NSF, Grant No.\
DMS-010-3998.}, and S.-T.\ Yau\thanks{Research supported in part
by the NSF, Grant No.\ 33-585-7510-2-30.}}
\date{March 2006}

\newtheorem{Def}{Def.}[section]
\newtheorem{Thm}[Def]{Theorem}
\newtheorem{Prp}[Def]{Proposition}
\newtheorem{Lemma}[Def]{Lemma}

\newtheorem{Corollary}[Def]{Corollary}
\newcommand{\Proof}{{\em{Proof. }}}
\newcommand{\QED}{\ \hfill $\FBox$ \\[1em]}
\newcommand{\spc}{\;\;\;\;\;\;\;\;\;\;}
\newcommand{\bra}{\mbox{$< \!\!$ \nolinebreak}}
\newcommand{\ket}{\mbox{\nolinebreak $>$}}
\newcommand{\lbra}{\langle}
\newcommand{\lket}{\rangle}
\newcommand{\C}{\mathbb{C}}
\newcommand{\R}{\mathbb{R}}
\newcommand{\1}{\mbox{\rm 1 \hspace{-1.05 em} 1}}
\newcommand{\Z}{\mathbb{Z}}
\newcommand{\sZ}{\mbox{\rm \bf \scriptsize Z}}

\newcommand{\N}{\mathbb{N}}
\newcommand{\sN}{\mbox{\rm \scriptsize I \hspace{-.8 em} N}}

\newcommand{\beq}{\begin{equation}}
\newcommand{\eeq}{\end{equation}}

\newcommand{\umax}{u_{\mbox{\tiny{max}}}}

\newcommand{\FBox}{\rule{2mm}{2.25mm}}

\newcommand{\LE}{{\mbox{\tiny{L}}}}
\newcommand{\Hp}{{\mbox{\tiny{H$+$}}}}
\newcommand{\Hm}{{\mbox{\tiny{H$-$}}}}
\newcommand{\Hpm}{{\mbox{\tiny{H$\pm$}}}}

\begin{document}
\maketitle
\begin{abstract}
We consider the Cauchy problem for the massless scalar wave equation
in the Kerr geometry for smooth initial data compactly supported outside the
event horizon. We prove that the solutions decay in time
in~$L^\infty_{\mbox{\tiny{loc}}}$. The proof is based on a
representation of the solution as an infinite sum over the angular
momentum modes, each of which is an integral of the energy
variable~$\omega$ on the real line. This integral representation
involves solutions of the radial and angular ODEs which arise in the
separation of variables.
\end{abstract}

\tableofcontents

\section{Introduction}
\setcounter{equation}{0} In this paper we study the long-time
dynamics of massless scalar waves in the Kerr geometry. We prove
that solutions of the Cauchy problem with smooth initial data which
is compactly supported outside the event horizon, decay
in~$L^\infty_{\mbox{\scriptsize{loc}}}$. Our starting point is the
integral representation for the propagator~\cite{FKSY}, which
involves an integral over a complex contour in the energy
variable~$\omega$. In order to study the long-time dynamics, we must
deform the contour to the real line. To this end, we carefully
analyze the solutions of the associated radial and angular ODEs
which arise in the separation of variables. In particular, we show
that the integrand in our representation has no poles on the real
axis. We call such poles {\em{radiant modes}}, because in a
dynamical situation they would lead to continuous radiation coming
out of the ergosphere.

We now set up some notation and state our main result. As in~\cite{FKSY},
we choose Boyer-Lindquist coordinates~$(t, r, \vartheta, \varphi)$ with
$r>0$, $0 \leq \vartheta \leq \pi$, $0 \leq \varphi < 2\pi$,
in which the Kerr metric takes the form
\beq ds^2 \;=\;
\frac{\Delta}{U} \:(dt \:-\: a \:\sin^2 \vartheta \:d\varphi)^2
\:-\: U \left( \frac{dr^2}{\Delta} + d\vartheta^2 \right) \:-\:
\frac{\sin^2 \vartheta}{U} \:(a \:dt \:-\: (r^2+a^2) \:d\varphi)^2
\label{eq:0}
\eeq
with
\[ U(r, \vartheta) \;=\; r^2 + a^2 \:\cos^2 \vartheta \;,\spc
\Delta(r) \;=\; r^2 - 2 M r + a^2  \; , \] where $M$ and $aM$
denote the mass and the angular momentum of the black hole,
respectively. We restrict attention to the
{\em{non-extreme case}} $M^2 > a^2$, where the function $\Delta$
has two distinct zeros,
\[ r_0 \;=\; M \:-\: \sqrt{M^2 - a^2} \spc {\mbox{and}} \spc r_1 \;=\; M \:+\: \sqrt{M^2 -
a^2} \; , \]
corresponding to the Cauchy and the event horizon,
respectively. We consider only the region $r>r_1$
outside the event horizon, and thus $\Delta>0$. The ergosphere
is the region where the Killing vector $\frac{\partial}{\partial t}$ is space-like,
that is where
\begin{equation}
r^{2}-2Mr+a^{2}\,\cos^{2}\vartheta < 0\:.
\end{equation}
The ergosphere lies outside the event
horizon $r=r_{1}$, and its boundary intersects the event
horizon at the poles $\vartheta=0,\,\pi$.
\begin{Thm} \label{thmmain}
Consider the Cauchy problem for the wave equation in the Kerr geometry for
smooth initial data which is compactly supported outside the event horizon and has
fixed angular momentum in the direction of the rotation axis of the black hole, i.e.\
for some~$k \in \Z$,
\[ (\Phi_0, \partial_t \Phi_0) \:=\: e^{-i k \varphi} \:(\Phi_0, \partial_t \Phi_0)(r, \vartheta)
\;\in\; C^\infty_0((r_1, \infty) \times S^2)^2\:. \]
Then the solution decays in~$L^\infty_{\mbox{\scriptsize{loc}}}((r_1, \infty) \times S^2)^2$
as~$t \to \infty$.
\end{Thm}

The study of linear hyperbolic equations in a black hole geometry has a long history.
Regge and Wheeler~\cite{RW} considered the radial equation for metric perturbations
of the Schwarzschild metric. In the late 1960s and early 1970s, Carter, Teukolsky and
Chandrasekhar discovered that the equations describing scalar, Dirac, Maxwell and
linearized gravitational fields in the Kerr geometry are separable
into ordinary differential equations (see~\cite{C}).
Much research has been done concerning the long-time behavior of the solutions of these
equations, through both numerical and analytical methods.
Price~\cite{Price} gave arguments which indicated decay of
solutions of the scalar wave equation in the Schwarzschild geometry.
Press and Teukolsky~\cite{PT} did a numerical study which strongly
suggested
the absence of unstable modes, and Whiting~\cite{W} later proved that for~$\omega$
in the complex plane, such unstable modes cannot exist.
This ``mode stability'' does not rule out that there might be
unstable modes for real~$\omega$ (what we call radiant modes).
Furthermore, mode stability does not lead to any statement on
the Cauchy problem. Finally,
Kay and Wald~\cite{KW} used energy estimates to prove a boundedness
result for solutions of the scalar wave equation in the Schwarzschild geometry.

Unfortunately, these energy methods cannot be used in a
rotating black hole geometry,
because the energy density is indefinite inside the ergosphere,
making it impossible to introduce a positive definite conserved
scalar product. This difficulty was dealt with in~\cite{FKSY, FS},
where Whiting's mode stability result was combined with estimates
for the resolvent and for the radial and angular ODEs.
In~\cite{FKSY} we established an integral representation
which expresses the solution as a contour integral of an integrand
involving the separated radial and angular eigenfunctions over a
contour staying within a neighborhood located arbitrarily close to
the real axis. This integral representation is the starting point
of the present paper. After deforming the contours onto the real axis,
we can prove decay using the Riemann-Lebesgue
Lemma, similar to the case of the Dirac equation~\cite{FKSY03}.
We remark that the decay result of our paper would not
be expected to hold for a massive scalar field satisfying the
Klein-Gordon equation, as indicated in~\cite{CY}.

Finally, we note that the problem considered here is closely
related to one of the major open questions in general relativity;
namely the problem of linearized stability of the Kerr metric.
For the stability under metric perturbations one considers the equation
for linearized gravitational waves, which can be identified with the general
wave equation for spin~$s=2$ (see~\cite{C}).
Thus replacing scalar waves ($s=0$) by gravitational
waves ($s=2$), the above theorem would prove linearized stability
of the Kerr metric. However, the analysis for~$s=2$ would be considerably more difficult
due to the complexity of the linearized Einstein equations.
Nevertheless, we regard this paper as a first step towards proving
linearized stability of the Kerr metric.

\section{Preliminaries}
\setcounter{equation}{0}
We recall a few constructions and results from~\cite{FKSY, FS} which will be
needed later on. As radial variable we usually work with
the Regge-Wheeler variable~$u \in \R$ defined by
\begin{equation} \label{51a}
\frac{du}{dr} \;=\; \frac{r^2+a^2}{\Delta} \;;
\end{equation}
then~$u=-\infty$ corresponds to the event horizon.
It is most convenient to write the wave equation in the
Hamiltonian form
\begin{equation}\label{Hamform}
i\,\partial_{t}\Psi=H\,\Psi,
\end{equation}
where~$\Psi = (\Phi, i \partial_t \Phi)$. The Hamiltonian can be written as
\beq \label{HAb}
H \;=\; \left( \begin{array}{cc} 0 & 1 \\ A & \beta \end{array}
\right) ,
\eeq
where
\begin{eqnarray}
A &=& \frac{1}{\rho} \left[-\frac{\partial}{\partial u} (r^2+a^2)
\frac{\partial}{\partial u} - \frac{\Delta}{r^2+a^2}\: \Delta_{S^2}
- \frac{a^2 k^2}{r^2+a^2} \right] \label{Adef} \\
\beta &=& -\frac{2 a k}{\rho} \left( 1 - \frac{\Delta}{r^2+a^2} \right) \label{bdef} \\
\rho &=& r^2+a^2 \:-\: a^2\: \sin^2 \vartheta\: \frac{\Delta}{r^2+a^2} \:. \label{rdef}
\end{eqnarray}
The operators~$A$ and~$\beta$ are symmetric on the Hilbert
space~$L^2(\R \times S^2, d\mu)^2$ with
the measure
\beq \label{mudef}
d\mu \;:=\; \rho\: du\: d\cos \vartheta \:.
\eeq
It is immediately verified that the Hamiltonian is symmetric
with respect to the bilinear form
\beq
\bra \Psi_1, \Psi_2 \ket \;=\; \int_{\R \times S^2}
\lbra \Psi_1, \left( \begin{array}{cc} A & 0 \\ 0 & 1 \end{array} \right)
\Psi_2 \lket_{\C^2} \:d\mu\:. \label{ESP}
\eeq
As is worked out in detail in~\cite{FKSY}, the inner
product~$\bra \Psi, \Psi \ket$ is the physical energy of~$\Psi$.
Therefore, we refer to~$\bra .,. \ket$ as the {\em{energy scalar product}}.
The fact that the energy scalar product is not positive definite can be
understood from the fact that the operator~$A$ is not positive
on~$L^2(\R \times S^2, d\mu)$.

Using the ansatz
\begin{equation}\label{separansatz}
\Phi(t,r,\vartheta,\varphi)=e^{-i\omega
t-ik\varphi}\:R(r)\:\Theta(\vartheta),
\end{equation}
the wave equation can be separated into an angular and a radial ODE,
\begin{equation} \label{odes}
{\cal{R}}_{\omega ,k}\,R_\lambda \;=\; -\lambda \, R_{\lambda},\qquad  {\cal{A}}_{\omega ,k}\,\Theta_\lambda \;=\; \lambda \, \Theta_\lambda\:.
\end{equation}
Here the angular operator~${\cal{A}}_{\omega, k}$ is also called the
{\em{spheroidal wave operator}}.  The separation constant~$\lambda$ is an
eigenvalue of~${\cal{A}}_{\omega, k}$ and can thus be regarded as
an angular quantum number. In~\cite{FS} it was shown that if~$\omega$
is in a small neighborhood of the real line, more precisely if
\[ \omega \;\in\; U_\varepsilon \;:=\; \left\{ \omega \in \C \:|\: |{\mbox{\rm{Im}}}\, \omega| <
\frac{\varepsilon}{1 + |{\mbox{\rm{Re}}}\, \omega|}\right\} , \]
then for sufficiently small~$\varepsilon>0$ the angular operator~${\cal{A}}_{\omega ,k}$
has a purely discrete spectrum~$(\lambda_n)_{n \in \N}$ with corresponding
one-dimensional eigenspaces which span the Hilbert space~$L^2(S^2)$.
We denote the projections onto the eigenspaces by~$Q_n(k,\omega)$.
These projections as well as the corresponding eigenvalues~$\lambda_n$
are holomorphic in~$\omega \in U_\varepsilon$.
In analogy to the eigenvalues~$l(l+1)$ of the Laplacian on the sphere,
the angular eigenvalues~$\lambda_n$ grow quadratically for large~$n$ in the sense
that there is a constant~$C(k, \omega)>0$ such that
\beq \label{quadra}
|\lambda_n(k, \omega)| \;\geq\; \frac{n^2}{C(k, \omega)} \spc {\mbox{for all~$n \in \N$}}\:.
\eeq
We set
\beq \label{o0def}
\omega_0 \;=\; -\frac{ak}{r_1^2+a^2}
\eeq
with~$r_1$ the event horizon and use the notation
\beq \label{Odef}
\Omega(\omega) \;=\; \omega - \omega_0\:.
\eeq

In order to bring the radial equation into a convenient form, we introduce
a new radial function~$\phi(r)$ by
\[ \phi(r) \;=\; \sqrt{r^2+a^2}\: R(r) \;. \]
Then in the Regge-Wheeler variable, the radial equation can be written
as the ``Schr{\"o}dinger equation''
\begin{equation}
\left(-\frac{d^2}{d u^2} + V(u) \right) \phi(u) \;=\; 0
\label{5ode}
\end{equation}
with the potential
\begin{equation} \label{5V}
V(u) \;=\; -\left( \omega + \frac{ak}{r^2+a^2} \right)^2 \:+\:
\frac{\lambda_n(\omega)\:\Delta}{(r^2+a^2)^2} \:+\: \frac{1}{\sqrt{r^2+a^2}}\; \partial_u^2 \sqrt{r^2+a^2} \:.
\end{equation}

In~\cite{FKSY} we derived an integral representation for the solution of the
Cauchy problem of the following form,
\begin{eqnarray}
\lefteqn{ \Psi(t,r,\vartheta, \varphi) } \nonumber \\
&=& -\frac{1}{2\pi i} \sum_{k \in \sZ} e^{-i k \varphi} \sum_{n \in \sN}\;
\lim_{\varepsilon \searrow 0} \left( \int_{C_\varepsilon} -\int_{\overline{C_\varepsilon}} \right) d\omega\; e^{-i \omega t}\:
(Q_{k,n}(\omega)\: S_{\infty}(\omega)\: \Psi_0^k)(r, \vartheta) \:.\quad \label{intrep}
\end{eqnarray}
Here the integration contour~$C_\varepsilon$ must lie inside the set~$U_\varepsilon$.

\section{Asymptotic Estimates for the Radial Equation}
\setcounter{equation}{0}
\subsection{Holomorphic Families of Radial Solutions} \label{secholo}
In this section we fix the angular quantum numbers~$k, n$ and
consider solutions~$\acute{\phi}$
and~$\grave{\phi}$ of the Schr{\"o}dinger equation~(\ref{5ode}) which satisfy
the following asymptotic boundary conditions
on the event horizon and at infinity, respectively,
\begin{eqnarray}
\lim_{u \to -\infty} e^{-i \Omega u} \:\acute{\phi}(u) &=& 1 \:,\spc
\lim_{u \to -\infty} \left(e^{-i \Omega u} \:\acute{\phi}(u) \right)' \;=\; 0  \label{abc1} \\
\lim_{u \to \infty} e^{i \omega u} \:\grave{\phi}(u) &=& 1 \:,\spc\;\;\;\;\,
\lim_{u \to \infty} \left(e^{i \omega u} \:\grave{\phi}(u) \right)' \;=\; 0\:.  \label{abc2}
\end{eqnarray}
These solutions were introduced in~\cite{FKSY} for~$\omega$ in the
lower complex half plane intersected with~$U_\varepsilon$. Here we will show that they are
holomorphic in~$\omega$, and we will extend their definition to a
larger $\omega$-domain. More precisely, we prove the following two
theorems.
\begin{Thm} \label{thmh1}
The solutions~$\acute{\phi}$ are well-defined on the domain
\[ D \;=\; U_\varepsilon \cap \left\{ \omega \in \C \:|\: {\mbox{\rm{Im}}}\, \omega \;\leq\;
\frac{r_1-r_0}{2 (r_1^2+a^2)} \right\} . \]
They form a holomorphic family of solutions in the sense that for every
fixed~$u \in \R$ and~$n \in \N$, the function
$\acute{\phi}(u)$ is holomorphic in~$\omega \in D$.
\end{Thm}

\begin{Thm} \label{thmh2} For every angular momentum number~$n$ there is an open
set~$E$ containing the real line except for the origin,
\beq \label{E0def}
E \;\supset\; E_0 \;:=\; U_\varepsilon \cap
\left\{ \omega \in \C \:|\: {\mbox{\rm{Im}}}\, \omega \leq 0
{\mbox{ and }} \omega \neq 0 \right\} ,
\eeq
such that the solutions~$\grave{\phi}$ are well-defined for all~$\omega \in E$ and
form a holomorphic family on~$E$.
\end{Thm}

For the proofs we will rewrite the Schr{\"o}dinger equation with boundary
conditions~(\ref{abc1}, \ref{abc2}) as an integral equation (which in different contexts is
called Lipman-Schwinger or Jost equation). Then we will perform a perturbation
expansion and get estimates for all the terms of the expansion. To introduce
the method, we begin with the solutions~$\acute{\phi}$;
the solutions~$\grave{\phi}$ will be treated later with a similar technique.
First we write the Schr{\"o}dinger equation~(\ref{5ode}) in the form
\beq \label{sch}
\left(-\frac{d^2}{du^2} - \Omega^2 \right) \:\acute{\phi}(u)
\;=\; -W(u) \:\acute{\phi}(u)
\eeq
with a potential~$W=\Omega^2 + V(u)$ which vanishes at~$u=-\infty$.
We define the {\em{Green's function}} of the differential
operator~$-\partial_u^2 - \Omega^2$ by the distributional equation
\beq \label{gv}
(-\partial_v^2-\Omega^2)\: S(u,v) \;=\; \delta(u-v)\:.
\eeq
The Green's function is not unique; we choose it such that its support
is contained in the region~$v \leq u$; i.e.
\beq \label{Sdef}
S(u,v) \;=\; \Theta(u-v)\:\times\: \left\{ \begin{array}{cl}
\displaystyle \frac{1}{2i \Omega} \left( e^{-i \Omega (u-v)}
- e^{i \Omega (u-v)} \right) & {\mbox{if~$\Omega \neq 0$}} \\
v-u & {\mbox{if~$\Omega = 0$}}\:. \end{array} \right.
\eeq
(Here~$\Theta$ denotes the Heaviside function defined by~$\Theta(x)=1$
if~$x \geq 0$ and~$\Theta(x)=0$ otherwise.)
We multiply~(\ref{sch}) by the Green's function and integrate,
\[ \int_{-\infty}^\infty S(u,v) \left((-\partial_v^2-\Omega^2)
\:(\acute{\phi}(v) - e^{i \Omega u}) \right) dv
\;=\; -\int_{-\infty}^\infty S(u,v)\: W(v)\: \acute{\phi}(v)\: dv\:. \]
If we assume for the moment that~$\acute{\phi}$ satisfies the desired
boundary conditions~(\ref{abc1}), we can integrate by parts on the left
and use~(\ref{gv}). This gives the Lipman-Schwinger equation
\[ \acute{\phi}(u) \;=\; e^{i \Omega u} - \int_{-\infty}^u S(u,v)\:
W(v)\: \acute{\phi}(v)\: dv\:, \]
which in the context of potential scattering is also called
{\em{Jost equation}} (see e.g.~\cite{AR}).
Its significance lies in the fact that we can now easily perform a
perturbation expansion in the potential~$W$. Namely, taking for~$\acute{\phi}$
the ansatz as the perturbation series
\beq \acute{\phi} \;=\; \sum_{l=0}^\infty \phi^{(l)}\:, \label{pe} \eeq
we are led to the iteration scheme
\beq \label{iter} \left.
\begin{array}{rcl}
\phi^{(0)}(u) &=& e^{i \Omega u} \\
\phi^{(l+1)}(u) &=& \displaystyle -\int_{-\infty}^u S(u,v)\:
W(v)\: \phi^{(l)}(v)\: dv \:. \end{array} \right\}
\eeq
This iteration scheme can be used for constructing solutions of the
Jost equation, and this will give us the functions~$\acute{\phi}$ with
the desired properties. \\[.5em]
{\em{Proof of Theorem~\ref{thmh1}.}}
Fix~$\omega \in D$. As the potential~$W$ is smooth in~$r$ and vanishes
on the event horizon, we know that~$W$ has near~$r_1$ the
asymptotics~$W={\mathcal{O}}(r-r_1)$. This means
in the Regge-Wheeler variable~(\ref{51a}) that~$W$ decays exponentially
as~$u \to -\infty$. More precisely, there is a constant~$c>0$ such that
\beq \label{Wbound}
|W(u)| \;\leq\; c\:e^{\gamma u} \spc {\mbox{with}} \spc
\gamma \::=\; \frac{r_1-r_0}{r_1^2+a^2}\:.
\eeq

Let us show inductively that
\beq \label{induc}
|\phi^{(l)}(u)| \;\leq\; \mu^l \:e^{- {\mbox{\scriptsize{\rm{Im}}}}\, \Omega\: u}
\spc {\mbox{with}} \spc
\mu \;:=\; \frac{c \:e^{\gamma u}}{(\gamma - {\mbox{\rm{Im}}}\, \Omega
- |{\mbox{\rm{Im}}}\, \Omega|)^2} \:.
\eeq
In the case~$l=0$, the claim is obvious from~(\ref{iter}). Thus assume that~(\ref{induc})
holds for a given~$l$. Then, estimating the integral equation in~(\ref{iter})
using~(\ref{Wbound}), we obtain
\beq \label{phiind}
|\phi^{(l+1)}(u)| \;\leq\; c \:\mu^l
\int_{-\infty}^u |S(u,v)| \:e^{(\gamma - {\mbox{\scriptsize{\rm{Im}}}}\, \Omega)\: v}\: dv \:.
\eeq
The Green's function~(\ref{Sdef}) can be estimated in the case~$v \leq u$ by
\begin{eqnarray*}
|S(u,v)| \;=\; \frac{u-v}{2} \left| \int_0^1 e^{-i \Omega (u-v)\:\tau} \:d\tau \right|
\;\leq\; (u-v)\: e^{|{\mbox{\scriptsize{\rm{Im}}}}\, \Omega|\, (u-v)} \:.
\end{eqnarray*}
Substituting this inequality in~(\ref{phiind}) gives
\[ |\phi^{(l+1)}(u)| \;\leq\; c \:\mu^l\:e^{|{\mbox{\scriptsize{\rm{Im}}}}\, \Omega|\, u}
\int_{-\infty}^u (u-v) \:e^{(\gamma - {\mbox{\scriptsize{\rm{Im}}}}\, \Omega
- |{\mbox{\scriptsize{\rm{Im}}}}\, \Omega|)\: v}\: dv \:. \]
Since the parameter~$\alpha := \gamma - {\mbox{{\rm{Im}}}}\, \Omega
- |{\mbox{{\rm{Im}}}}\, \Omega|$ is positive according to the definition
of~$D$, we can carry out the integral as follows,
\[ \int_{-\infty}^u (u-v)\: e^{\alpha v}\: dv \;=\;
\left(u - \frac{d}{d\alpha} \right) \int_{-\infty}^u e^{\alpha v}\: dv
\;=\; \left(u - \frac{d}{d\alpha} \right) \frac{e^{\alpha u}}{\alpha}
\;=\; \frac{e^{\alpha u}}{\alpha^2}\:. \]
This gives~(\ref{induc}) with~$l$ replaced by~$l+1$.

Since for~$u$ on a compact interval, the
analytic dependence of the solutions in~$\omega$ from the coefficients
and the initial conditions follows immediately from the Picard-Lindel{\"o}f
Theorem, it suffices to consider the region~$u<u_0$ for any~$u_0 \in \R$.
By choosing~$u_0$ sufficiently small, we can arrange that~$\mu<1/2$ for
all~$u < u_0$. Then the estimate~(\ref{induc}) shows that the perturbation
series~(\ref{pe}) converges absolutely, uniformly in~$u \in (-\infty, u_0)$.
Using similar estimates for the $u$-derivatives of~$\phi^{(l)}$, one sees
furthermore that the perturbation series~(\ref{induc}) can be differentiated
term by term, and using~(\ref{gv}) we find that~$\acute{\phi}$ is indeed a
solution of~(\ref{sch}). Furthermore,
\[ \acute{\phi}(u) - e^{i \Omega u} \;=\; \sum_{l=1}^\infty \phi^{(l)}(u)\:, \]
and taking the limit~$u \to \infty$ and using~(\ref{induc}) we find that
the right side goes to zero. Using the same argument for the first derivatives,
we obtain~(\ref{abc1}).

In order to prove that~$\acute{\phi}$ is analytic in~$\omega$, we first note
that if~$\Omega \neq 0$, we can differentiate the perturbation series~(\ref{pe})
term by term and verify that the Cauchy-Riemann equations are satisfied
(note that~$\lambda_n$ is holomorphic in~$\omega$ according to~\cite{FS}).
Since~$\acute{\phi}$ is bounded near~$\Omega=0$, it is also analytic
at~$\Omega=0$.
\QED

We turn to the solutions~$\grave{\phi}$. In analogy to~(\ref{sch}), we now
write the Schr{\"o}dinger equation as
\beq \label{sch2}
\left(-\frac{d^2}{du^2} - \omega^2 \right) \:\grave{\phi}(u)
\;=\; -W(u) \:\grave{\phi}(u)
\eeq
with
\beq \label{Wdef2}
W(u) \;=\; -\omega \frac{2 ak}{r^2+a^2} \:-\: \frac{(ak)^2}{(r^2+a^2)^2} \:+\:
\frac{\lambda_n\:\Delta}{(r^2+a^2)^2} \:+\: \frac{1}{\sqrt{r^2+a^2}}\; \partial_u^2 \sqrt{r^2+a^2} \:.
\eeq
Assuming that~$\omega \neq 0$, we choose the Green's function as
\beq \label{Sdef2}
S(u,v) \;=\; \frac{1}{2i \omega} \left( e^{-i \omega (v-u)}
- e^{i \omega (v-u)} \right)\: \Theta(v-u)\:.
\eeq
The corresponding Jost equation is
\[ \grave{\phi}(u) \;=\; e^{-i \omega u} - \int_{u}^\infty S(u,v)\:
W(v)\: \grave{\phi}(v)\: dv\:. \]
The perturbation series ansatz
\beq \grave{\phi} \;=\; \sum_{l=0}^\infty \phi^{(l)} \label{pe2} \eeq
leads to the iteration scheme
\beq \label{iter2} \left.
\begin{array}{rcl}
\phi^{(0)}(u) &=& e^{-i \omega u} \\
\phi^{(l+1)}(u) &=& \displaystyle -\int_u^\infty S(u,v)\:
W(v)\: \phi^{(l)}(v)\: dv \:. \end{array} \right\}
\eeq
Note that, in contrast to the exponential decay~(\ref{Wbound}),
now the potential~$W$, (\ref{Wdef2}), has only polynomial decay.
As a consequence, the iteration scheme allows us to construct~$\grave{\phi}$
only inside the set~$E_0$ as defined in~(\ref{E0def}).
\begin{Lemma} \label{lemmah2}
The solutions~$\grave{\phi}$ are well-defined for every~$\omega \in E_0$.
They form a holomorphic family in the interior of~$E_0$.
\end{Lemma}
{\Proof}
Fix~$\omega \in E_0$. Then~$\omega \neq 0$ and~${\mbox{Im}}\, \omega \leq 0$,
and this allows us to estimate the potential~(\ref{Wdef2}) and the Green's
function~(\ref{Sdef2}) for~$u,v > u_0$ and some~$u_0>0$ by
\beq \label{WSest}
|W(v)| \;\leq\; \frac{c}{v^2} \:,\spc
|S(u,v)| \;\leq\; \frac{1}{|\omega|}\: e^{{\mbox{\scriptsize{\rm{Im}}}}\, \omega
\: (u-v)}\:.
\eeq
Let us show by induction that
\[ |\phi^{(l)}(u)| \;\leq\; \frac{1}{l!} \left( \frac{c}{|\omega|\: u} \right)^l
e^{{\mbox{\scriptsize{\rm{Im}}}}\, \omega\: u} \:. \]
For~$l=0$ this is obvious from~(\ref{iter2}), whereas the induction step
follows by estimating the integral equation in~(\ref{iter2}) with~(\ref{WSest}),
\begin{eqnarray*}
|\phi^{(l+1)}(u)| &\leq& \frac{1}{l!} \left( \frac{c}{|\omega|} \right)^n
\int_u^\infty \frac{1}{|\omega|}\: e^{{\mbox{\scriptsize{\rm{Im}}}}\, \omega
\: (u-v)}\: \frac{c}{v^{2+l}}\: e^{{\mbox{\scriptsize{\rm{Im}}}}\, \omega\: v} \\
&=& \frac{1}{(l+1)!} \left( \frac{c}{|\omega|\: u} \right)^{l+1}
e^{{\mbox{\scriptsize{\rm{Im}}}}\, \omega\: u} \:.
\end{eqnarray*}
Hence the perturbation series~(\ref{pe2}) converges absolutely, locally uniformly
in~$u$. It is straightforward to check that~$\grave{\phi}$ satisfies the
Schr{\"o}dinger equation~(\ref{sch2}) with the correct boundary values~(\ref{abc2}).
If~${\mbox{Im}}\, \omega <0$, one can differentiate the series~(\ref{pe2}) term by term
with respect to~$\omega$ and verify that the Cauchy-Riemann equations are satisfied.
\QED

It remains to analytically extend the solutions~$\grave{\phi}$ for fixed~$n$ to a
neighborhood of any point~$\omega_0 \in \R \setminus \{0\}$. To this end, we need
good estimates of the derivatives of~$\grave{\phi}$ with respect to~$\omega$ and~$u$.
It is most convenient to work with the functions
\beq \label{pe3}
\psi^{(l)}(u) \;:=\; (2 i \omega)^l \:e^{i \omega u}\: \phi^{(l)}(u) \:,
\eeq
for which the iteration scheme~(\ref{iter2}) can be written as
$\psi^{(0)} =1$ and
\beq \label{iter3}
\psi^{(l+1)}(u) \;=\; \int_u^\infty (e^{-2 i \omega (v-u)} - 1)\:
W(v)\: \psi^{(l)}(v)\: dv \:.
\eeq
\begin{Lemma} \label{lemmacombi}
For every~$\omega_0 \in \R \setminus \{0\}$ and~$n \in \N$,
there are positive constants~$c, K, \delta$, such that
for all~$\omega \in E_0 \cap B_\delta(\omega_0)$ with
${\mbox{\rm{Im}}}\, \omega < 0$ and all~$p, q, n \in \N$
the following inequality holds,
\beq \label{ineq}
\left| \left(\frac{\partial}{\partial \omega} \right)^p
\left( \frac{\partial}{\partial u} \right)^q \psi^{(l)}(u) \right| \;\leq\;
c^{1+l+p}\:K^q\:\frac{p!\:q!}{l!}\: \frac{1}{u^{l+q}}\:.
\eeq
\end{Lemma}
{\Proof} According to~\cite{FS}, $\lambda_n$ is holomorphic in a neighborhood
of~$\omega_0$, and thus (for example using the Cauchy integral formula)
its derivatives can be bounded in~$B_\delta(\omega_0)$ by
\[ |\partial_\omega^p \lambda_n(\omega)| \;\leq\;
\left(\frac{K}{2} \right)^{1+p} p! \]
for suitable~$K>0$. Since the potential~$W$, (\ref{Wdef2}), is also
holomorphic in~$r$ (in a suitable neighborhood of the positive real axis) and has
quadratic decay, its derivatives can be estimated by
\beq \label{West}
|\partial_\omega^p \partial_u^q W(u)| \;\leq\;
\left( \frac{K}{2} \right)^{1+p+q} \frac{p!\:q!}{u^{2+q}} \:.
\eeq
We choose~$c$ so large that the following conditions hold,
\beq \label{condc}
c \;>\; 16\,K \:,\spc \frac{K\:e^{\frac{1}{K}}}{(\omega_0-\delta)\: c} \;\leq\; \frac{1}{2} \:.
\eeq

We proceed to prove~(\ref{ineq})
by induction in~$l$. For~$l=0$ there is nothing to prove.
Thus assume that~(\ref{ineq}) holds for a given~$l$.
Using the induction hypothesis together with~(\ref{West}), we can then
estimate the derivatives of the product~$W \psi^{(l)}$ as follows,
\begin{eqnarray*}
|\partial_\omega^p \partial_u^q (W \psi^{(l)})| &\leq&
\sum_{a=0}^p \left( \!\!\begin{array}{c} p \\ a \end{array} \!\!\right)
\sum_{b=0}^q \left( \!\!\begin{array}{c} q \\ b \end{array} \!\!\right)
\left(\frac{K}{2} \right)^{1+a+b}\: \frac{a!\:b!}{u^{2+b}} \;
\frac{c^{1+l+p-a}\:K^{q-b}}{u^{l+q-b}} \:\frac{(p-a)!\:(q-b)!}{l!} \\
&=& \frac{c^{1+l+p}\:K^{1+q}}{u^{2+l+q}}\: \frac{p!\: q!}{l!}
\sum_{a=0}^p \left(\frac{K}{2c}\right)^a \sum_{b=0}^q \left(\frac{1}{2}\right)^b \:.
\end{eqnarray*}
According to~(\ref{condc}), the two remaining sums can be bounded by
the geometric series $\sum_{m=0}^\infty 2^{-m} = 2$, and thus
\beq \label{Wpsiest}
|\partial_\omega^p \partial_u^q (W \psi^{(l)})| \;\leq\;
4 \:\frac{c^{1+l+p}\:K^{1+q}}{u^{2+l+q}}\: \frac{p!\: q!}{l!}\:.
\eeq

Next we differentiate the integral equation~(\ref{iter3}),
\[ \partial^p_\omega \partial^q_u \psi^{(l+1)}(u) \;=\;
\sum_{r=0}^p \left( \!\!\begin{array}{c} p \\ r \end{array} \!\!\right)
\int_{-\infty}^\infty
\partial^r_\omega \partial^q_u \left[\Theta(v-u)\:(e^{-2 i \omega (v-u)} - 1) \right] \:
\partial^{p-r}_\omega \left( W \psi^{(l)}(v) \right)\:dv \]
(note that, since ${\mbox{\rm{Im}}}\, \omega < 0$, the
factor~$e^{-2 i \omega v}$ gives an exponential decay of the
integrand as $v \rightarrow \infty$).
After manipulating the partial derivatives as follows,
\[ \partial^r_\omega \partial^q_u \left[\Theta(v-u)\:(e^{-2 i \omega (v-u)} - 1) \right]
\;=\; (-\partial_v)^q
\left[\Theta(v-u) \left(\frac{v-u}{\omega} \: \partial_v \right)^r
(e^{-2 i \omega (v-u)} - 1) \right] , \]
the resulting $v$-derivatives can all be integrated by parts.
The boundary terms drop out, and we obtain
\[ \partial^p_\omega \partial^q_u \psi^{(l+1)}(u) \;=\;
\sum_{r=0}^p \left( \!\!\begin{array}{c} p \\ r \end{array} \!\!\right)
\int_u^\infty (e^{-2 i \omega (v-u)} - 1)
\left( \partial_v \:\frac{u-v}{\omega} \right)^r
\partial^q_v \partial^{p-r}_\omega \left( W \psi^{(l)}(v) \right) dv\:. \]
Since~$\omega$ is in the lower half plane,
we have the inequality~$|e^{-2 i \omega (v-u)}| \leq 1$. We conclude that
\beq \label{eqcombi}
\left| \partial^p_\omega \partial^q_u \psi^{(l+1)}(u) \right|
\;\leq\; 2 \sum_{r=0}^p \left( \!\!\begin{array}{c} p \\ r \end{array} \!\!\right)
\int_u^\infty \left| \left\{ \partial_v \:\frac{u-v}{\omega} \right\}^r
\partial^q_v \partial^{p-r}_\omega \left( W \psi^{(l)}(v) \right) \right| dv\:.
\eeq
The~$v$-derivatives in the curly brackets can act either on one of the
factors~$(u-v)$ or on the function~$W \psi^{(l)}$.
Taking into account the combinatorics, we obtain
\[ \left| \partial^p_\omega \partial^q_u \psi^{(l+1)}(u) \right|
\;\leq\; 2 \sum_{r=0}^p \left( \!\!\begin{array}{c} p \\ r \end{array} \!\!\right)
\frac{1}{\omega^r} \sum_{s=0}^r
\left( \!\!\begin{array}{c} r \\ s \end{array} \!\!\right)
r^{r-s} \int_u^\infty (v-u)^s
\left| \partial^{p-r}_\omega \partial^{q+s}_v
  \left( W \psi^{(l)}\right) \right| dv\:. \]
Using~(\ref{Wpsiest}), we get
\[ \left| \partial^p_\omega \partial^q_u \psi^{(l+1)}(u) \right|
\;\leq\; 8 \sum_{r=0}^p \sum_{s=0}^r
\frac{p!\:(q+s)!}{s!\:(r-s)!\:l!}\;\omega^{-r}\:r^{r-s}\:
c^{1+l+p-r}\:K^{1+q+s}
\int_u^\infty \frac{(v-u)^s}{v^{2+l+q+s}}\: dv\:. \]
Introducing the new variable~$\tau = \frac{u}{v}$, the integral can be
computed with iterative integrations by parts,
\begin{eqnarray*}
\lefteqn{ \int_u^\infty \frac{(v-u)^s}{v^{2+l+q+s}}\: dv \;=\;
\frac{1}{u^{1+l+q}} \int_0^1 (1-\tau)^s\: \tau^{l+q}\:d\tau } \\
&=& \frac{1}{u^{1+l+q}} \:\frac{(l+q)!}{(l+q+s)!} \int_0^1 (1-\tau)^s\:
\frac{d^s}{d\tau^s} \tau^{l+q+s}\:d\tau \\
&=& \frac{1}{u^{1+l+q}} \:\frac{(l+q)!\:s!}{(l+q+s)!} \int_0^1 \tau^{l+q+s}\:d\tau
\;=\; \frac{1}{u^{1+l+q}} \:\frac{(l+q)!\:s!}{(1+l+q+s)!}\:.
\end{eqnarray*}
We thus obtain
\[ \left| \partial^p_\omega \partial^q_u \psi^{(l+1)}(u) \right|
\;\leq\; 8 \: \frac{c^{1+l+p}\:K^{1+q}}{u^{1+l+q}}
\sum_{r=0}^p \sum_{s=0}^r
\frac{p!\:(q+s)!}{(r-s)!\:l!}\:\frac{K^{s}\:r^{r-s}}{(\omega c)^r}
\:\frac{(l+q)!}{(1+l+q+s)!}\:. \]
Using the elementary estimate
\[ \frac{(q+s)!\:(l+q)!}{(1+l+q+s)!}
\;=\; \frac{q!}{q+l+1}\cdot\frac{q+1}{q+l+2} \cdots
\frac{q+s}{q+l+s+1} \;\leq\; \frac{q!}{l+1}\:, \]
we obtain
\[ \left| \partial^p_\omega \partial^q_u \psi^{(l+1)}(u) \right|
\;\leq\; 8 \: \frac{c^{1+l+p}\:K^{1+q}}{u^{1+l+q}}\:
\frac{p!\:q!}{(l+1)!}
\sum_{r=0}^p \left( \frac{K}{\omega c} \right)^r \sum_{s=0}^r
\frac{1}{(r-s)!} \left(\frac{r}{K} \right)^{r-s} \:. \]
The last sum can be estimated by an exponential,
\[ \sum_{s=0}^r \frac{1}{(r-s)!} \left(\frac{r}{K} \right)^{r-s} \;\leq\;
\sum_{a=0}^r \frac{1}{a!} \left(\frac{r}{K} \right)^a
\;\leq\; \sum_{a=0}^\infty \frac{1}{a!} \left(\frac{r}{K} \right)^a \;=\;
\exp \left(\frac{r}{K} \right) . \]
According to~(\ref{condc}), we can now estimate the remaining sum over~$r$
by a geometric series,
\[ \sum_{r=0}^p \left( \frac{K}{\omega c} \right)^r \exp \left(\frac{r}{K} \right)
\;\leq\;
\sum_{r=0}^\infty \left( \frac{K\: e^{\frac{1}{K}}}{\omega c} \right)^r
\;\leq\; 2\:. \]
We thus obtain
\[ \left| \partial^p_\omega \partial^q_u \psi^{(l+1)}(u) \right|
\;\leq\; 16\: \frac{c^{1+l+p}\:K^{1+q}}{u^{1+l+q}}\:
\frac{p!\:q!}{(l+1)!} \;\leq\; \frac{c^{2+l+p}\:K^q}{u^{1+l+q}}\:
\frac{p!\:q!}{(l+1)!} \:, \]
where in the last step we again used~(\ref{condc}).
\QED

\noindent
{\em{Proof of Theorem~\ref{thmh2}.}}
According to~(\ref{pe2}, \ref{pe3}),
\[ \grave{\phi}(\omega, u) \;=\; e^{-i \omega u} \sum_{l=0}^\infty
\frac{1}{(2 i \omega)^l}\: \psi^{(l)}(\omega, u) \:. \]
Expanding~$\psi^{(l)}$ in a Taylor series in~$\omega$, we obtain the
formal expansion
\[ \grave{\phi}(\omega+\zeta, u) \;=\; e^{-i \omega u} \sum_{l=0}^\infty
\frac{1}{(2 i (\omega+\zeta))^l}\:
\sum_{p=0}^\infty \frac{\zeta^p}{p!}\:\partial_\omega^p \psi^{(l)}(\omega, u) \:. \]
Lemma~\ref{lemmacombi} allows us to estimate this expansion
for every~$\omega \in E_0 \cap B_\delta(\omega_0)$ with ${\mbox{\rm{Im}}}\, \omega < 0$ as follows,
\[ |\grave{\phi}(\omega+\zeta, u)| \;\leq\;
c \sum_{l=0}^\infty \frac{1}{l!} \left(\frac{c}{|\omega+\zeta|\: u}\right)^l\:
\sum_{p=0}^\infty (c |\zeta|)^p\:. \]
This expansion converges uniformly for $|\zeta|<\frac{c}{2}$.
Similarly, one can show that the series of $\zeta$-derivatives also converge
uniformly. Hence we can interchange differentiation with summation,
and a straightforward calculation shows that the
Cauchy-Riemann equations are satisfied. Thus the above expansion allows us
to extend~$\grave{\phi}$ analytically to the ball~$|\zeta|<\frac{c}{2}$.
Since the constant~$c$ is independent of~${\mbox{\rm{Im}}}\, \omega$,
we thus obtain an analytic extension of~$\grave{\phi}$ across the real line.
\QED

\subsection{A Continuous Family of Solutions near~$\omega=0$}
In Theorem~\ref{thmh2} we made no statement about the behavior of the
fundamental solutions~$\grave{\phi}$ at~$\omega=0$. Indeed, we cannot expect
the solutions to have a holomorphic extension in a neighborhood of~$\omega=0$.
But at least, after suitable rescaling, these solutions have a well-defined limit at~$\omega=0$:

\begin{Thm} \label{thmc1} For every angular momentum number~$n$, there is a
real solution~$\phi_0$ of the Schr{\"o}dinger equation~(\ref{5ode})
for~$\omega=0$ with the asymptotics
\beq \label{casy}
\lim_{u \to \infty} \: u^{\mu - \frac{1}{2}}\: \phi_0(u) \;=\; \frac{\Gamma(\mu)}{\sqrt{\pi}}
\spc {\mbox{with}}\spc \mu \;:=\; \sqrt{\lambda_n(0) + \frac{1}{4}}\:.
\eeq
This solution can be obtained as a limit of the solutions from Theorem~\ref{thmh2},
in the sense that for all~$u \in \R$,
\[ \phi_0(u) \;=\; \lim_{E_0 \ni \omega \to 0} \omega^{\mu-\frac{1}{2}}
\:\grave{\phi}(u) \qquad {\mbox{and}} \qquad
\phi'_0(u) \;=\; \lim_{E_0 \ni \omega \to 0} \omega^{\mu-\frac{1}{2}}
\:\grave{\phi}'(u) \:. \]
\end{Thm}
Note that the~$\lambda_n$ are the eigenvalues of the Laplacian on the sphere. They are clearly non-negative, and thus the parameter~$\mu$ in~(\ref{casy})
is positive.

Unfortunately, the function~$\phi_0$ cannot be constructed with the iteration
scheme~(\ref{iter2}) because if we put in the Green's function
for~$\omega=0$ (which is obtained from~(\ref{Sdef2}) by taking the
limit~$\omega \to 0$), we get the for~$\phi^{(1)}$ the
equation
\[ \phi^{(1)}(u) \;=\; \int_u^\infty (v-u)\:
W(v) \: dv \:, \]
and since~$W$ decays at infinity only quadratically, the integral diverges.
To overcome this problem, we combine the quadratically decaying part of the potential
with the unperturbed operator. More precisely, for any~$\omega$ in the set
\[ F \;:=\; \left\{ \omega \in \C \:|\: {\mbox{\rm{Im}}}\, \omega \leq 0 {\mbox{ and }}
|\omega| \leq (16 a k)^{-1} \right\} \:, \]
we write the Schr{\"o}dinger equation as
\[ \left(-\frac{d^2}{d u^2} + \frac{\mu^2-\frac{1}{4}}{u^2} - \omega^2 \right) \phi(u) \;=\; -W(u)\: \phi(u)\:, \]
where~$\mu(\omega) = (\lambda_n(\omega) - 2ak \omega + \frac{1}{4})^\frac{1}{2}$.
The potential~$W$ is continuous in~$\omega$ and bounded by
\beq \label{Wbound2}
|W(u)| \;\leq\; \frac{c}{u^3} \spc {\mbox{for all~$\omega \in F$}}\:.
\eeq
The solutions of the unperturbed Schr{\"o}dinger equation can be expressed with Bessel functions,
\[ h_1(u) \;=\; \sqrt{\frac{\pi u \omega}{2}} \: J_\mu(\omega u) \:, \qquad
h_2(u) \;=\; \sqrt{\frac{\pi u \omega}{2}} \: Y_\mu(\omega u) \:. \]
They have the following asymptotics,
\[ \left\{ \begin{array}{rcllrcll}
h_1(u) &\!\!\sim\!\!& \cos(\omega u) &\!\!\!,\;\; & h_2(u) &\!\!\sim\!\!& \sin(\omega u)
&\spc {\mbox{if~$\omega u \gg 1$}} \\
h_1(u) &\!\!\sim\!\!& \displaystyle \frac{\sqrt{\pi}\:\omega^{\mu+\frac{1}{2}}}{\Gamma(\mu+1)\: 2^{\mu+\frac{1}{2}}}\: u^{\mu+\frac{1}{2}} &\!\!\!,&
h_2(u) &\!\!\sim\!\!& \displaystyle \frac{\Gamma(\mu)\:2^{\mu-\frac{1}{2}}}{\sqrt{\pi}\:\omega^{\mu+\frac{1}{2}}} \:u^{-\mu+\frac{1}{2}}
&\spc {\mbox{if~$\omega u \ll 1$}}.
\end{array} \right. \]
The Green's function can be expressed in terms of the two fundamental solutions by the standard formula
\[ S(u,v) \;=\; \Theta(v-u)\: \frac{h_1(u)\: h_2(v) - h_1(v)\: h_2(u)}{w(h_1, h_2)} \:, \]
where~$w(h_1, h_2) = h_1' h_2 - h_1 h_2' = -\omega$ is the Wronskian.
The perturbation series ansatz
\beq \label{phidef}
\phi \;=\; \sum_{l=1}^\infty \phi^{(l)}
\eeq
now leads to the integral equation
\beq \label{it4}
\phi^{(l+1)}(u) \;=\; \int_u^\infty S(u,v)\:
W(v)\: \phi^{(l)}(v)\: dv \:.
\eeq
We choose the function~$\phi^{(0)}$ such that its asymptotics at
infinity is a multiple times the plane wave~$e^{-i \omega u}$, whereas
for~$\omega=0$, it has the asymptotics~(\ref{casy}),
\beq \label{it5}
\phi^{(0)}(u) \;=\; \omega^\mu\:(h_1 - i h_2)(u) \:.
\eeq

\begin{Lemma} \label{lemma32}
For any fixed~$n$ there is~$u_0 \in \R$ such that the iteration
scheme~(\ref{it5}, \ref{it4}) converges uniformly
for all~$u>u_0$ and~$\omega \in F$.
The functions~$\phi$ defined by~(\ref{phidef}) are solutions of the
Schr{\"o}dinger equation~(\ref{5ode}) with the asymptotics
\[ \left| \frac{\phi(u)}{\phi^{(0)}(u)} - 1 \right| \;\leq\;
\frac{c}{u} \]
and a constant~$c=c(n)$.
\end{Lemma}
{\Proof}
Using the asymptotic formulas for the Bessel functions, one sees
(similar to the estimate~\cite[eqn.~(4.4)]{AR} for $\mu=l+\frac{1}{2}$ and integer~$l$)
that for all~$v\geq u$ and~$\omega \in F$, the Green's function is bounded by
\beq \label{bGreen}
|S(u,v)| \;\leq\; C \: e^{{\mbox{\scriptsize{\rm{Im}}}}\,\omega\: (v-u)}
\left(\frac{u}{1+|\omega|\,u} \right)^{-\mu+\frac{1}{2}}
\left(\frac{v}{1+|\omega|\,v} \right)^{\mu+\frac{1}{2}}\:.
\eeq
Similarly, we can bound the Bessel functions in~(\ref{it5}) to get
\beq \label{bfkt}
\frac{1}{C} \;\leq\;
|\phi^{(0)}|\: e^{-{\mbox{\scriptsize{\rm{Im}}}}\,\omega u}
\left(\frac{u}{1+|\omega|\,u} \right)^{\mu-\frac{1}{2}} \;\leq\; C  \:.
\eeq

Let us show inductively that
\beq \label{induc4}
|\phi^{(l)}| \;\leq\; C\:e^{{\mbox{\scriptsize{\rm{Im}}}}\,\omega\,u}
\left(\frac{u}{1+|\omega|\,u} \right)^{-\mu+\frac{1}{2}} \left(\frac{Cc}{u} \right)^l \:.
\eeq
For~$l=0$ there is nothing to prove. The induction step follows from~(\ref{it4},
\ref{Wbound2}, \ref{bGreen})
\begin{eqnarray*}
|\phi^{(l+1)}| &\leq&
C \: e^{-{\mbox{\scriptsize{\rm{Im}}}}\,\omega u}
\left(\frac{u}{1+|\omega|\,u} \right)^{-\mu+\frac{1}{2}}
\int_u^\infty e^{2\,{\mbox{\scriptsize{\rm{Im}}}}\,\omega v}
\left(\frac{v}{1+|\omega|\,v} \right)
\frac{cC}{v^3}\:\left(\frac{Cc}{v} \right)^l dv \\
&\leq& C \: e^{{\mbox{\scriptsize{\rm{Im}}}}\,\omega u}
\left(\frac{u}{1+|\omega|\,u} \right)^{-\mu+\frac{1}{2}}
\left(\frac{Cc}{u} \right)^l \int_u^\infty
\frac{cC}{v^2}\: dv
\end{eqnarray*}

The lemma now follows immediately from~(\ref{induc4}, \ref{bfkt})
and by differentiating the series~(\ref{phidef}) with respect to~$u$.
\QED
{\em{Proof of Theorem~\ref{thmc1}.}}
>>From the asymptotics at infinity, it is clear that
\[ \phi \;=\; \left\{ \begin{array}{cl}
\omega^\mu\:\grave{\phi}
& {\mbox{if $\omega \neq 0$}} \\
\phi_0 & {\mbox{if $\omega = 0$}}\:. \end{array} \right. \]
Denoting the~$\omega$-dependence of~$\phi$ by a subscript, we thus need to prove that
for all~$u \in \R$,
\beq \label{task}
\lim_{F \ni \omega \to 0} \phi_\omega(u) \;=\; \phi_0(u) \:, \spc
\lim_{F \ni \omega \to 0} \phi'_\omega(u) \;=\; \phi'_0(u)\:.
\eeq
To simplify the problem, we first note that for~$u$ on a compact intervals, the
continuous dependence on~$\omega$ follows immediately from the Picard-Lindel{\"o}f
Theorem (i.e.\ the continuous dependence of solutions of ODEs on the coefficients
and initial values). Thus it suffices to prove~(\ref{task}) for large~$u$.
Furthermore, writing the Schr{\"o}dinger equation as
\[ (\partial_u - i \omega)(\partial_u + i \omega)\:\phi_\omega \;=\; -U\: \phi \:, \]
the potential~$U$ has quadratic decay at infinity. Thus, after the substitution
$(\partial_u - i \omega) = e^{i \omega u} \partial_u e^{-i \omega u}$,
we can multiply the above equation by~$e^{-i \omega u}$ and integrate to
obtain
\[ e^{-i \omega u}\: (\partial_u+i \omega)\:\phi_\omega(u)
\;=\; \int_u^\infty e^{-i \omega v}\:U(v)\:\phi_\omega(v)\:dv\:. \]
Here we emphasized the $\omega$-dependence by a subscript; note also
that the integral is well-defined in view of the asymptotics of~$\phi_\omega$
at infinity.
This equation shows that~$\phi'_\omega$ converges pointwise once
we know that~$\phi_\omega(u)$ converges uniformly in~$u$.
Hence it remains to show that for every~$\epsilon>0$ there
is~$u_0$ and~$\delta>0$ such that for all~$\omega \in F$ with $|\omega|<\delta$,
\beq \label{task2}
|\phi_\omega(u) - \phi_0(u)| \;<\; \varepsilon
\spc {\mbox{for all~$u>u_0$}} .
\eeq

To prove~(\ref{task2}) we use the uniform convergence of the
functions~$\phi_\omega^{(0)}$, (\ref{it5}), to choose~$\delta$ such that
for all~$\omega \in F$ with $|\omega|<\delta$,
\[ |\phi^{(0)}_\omega(u) - \phi^{(0)}_0(u)| \;<\; \frac{\varepsilon}{3}
\spc {\mbox{for all~$u>u_0$}} . \]
According to Lemma~\ref{lemma32}, we can by choosing~$u_0$ sufficiently large
arrange that
\[ |\phi^{(0)}_\omega(u) - \phi_\omega(u)| \;<\; \frac{\varepsilon}{3}
\spc {\mbox{for all~$u>u_0$ and~$\omega \in F$}} . \]
Now~(\ref{task2}) follows immediately from the estimate
\[ |\phi_\omega - \phi_0| \;\leq\; |\phi_\omega - \phi^{(0)}_\omega|
+ |\phi^{(0)}_\omega - \phi^{(0)}_0| + |\phi^{(0)}_0 - \phi_0| \:. \]

\vspace*{-.7cm}
\QED

\section{Global Estimates for the Radial Equation} \label{secode}
\setcounter{equation}{0}
Let $Y_1$ and $Y_2$ be two real fundamental solutions of the Schr{\"o}dinger
equation~(\ref{5ode}) for a general real and smooth potential~$V$.
Then their Wronskian
\beq \label{wronski}
w \;:=\; Y_1'(u)\: Y_2(u) - Y_1(u) \: Y_2'(u)
\eeq
is a constant. By flipping the sign of~$Y_2$, we can always arrange
that $w < 0$. We combine the
two real solutions into the complex function
\[ z \;=\; Y_1 + i Y_2 \;, \]
and denote its polar decomposition by
\begin{equation} \label{5g}
z \;=\; \rho\: e^{i \varphi}
\end{equation}
with real functions $\rho(u) \geq 0$ and $\varphi(u)$.
By linearity, $z$ is a solution of the complex Schr{{\"o}}dinger equation
\begin{equation} \label{5csch}
z'' \;=\; V \: z \;.
\end{equation}
Note that $z$ has no zeros because at every $u$ at least one of the
fundamental solutions does not vanish.

\subsection{The Complex Riccati Equation}
We introduce the function~$y$ by
\begin{equation}
y \;=\; \frac{z'}{z} \:. \label{5ydef}
\end{equation}
Since~$z$ has no zeros, the function~$y$ is smooth.
Moreover, it satisfies the complex Riccati equation
\begin{equation}
y' + y^2 \;=\; V \;. \label{5c}
\end{equation}
The fact that the solutions of the complex Riccati equation are
smooth will be helpful for getting estimates.
Conversely, from a solution of the Riccati equation one obtains the
corresponding solution of the Schr{\"o}dinger equation by integration,
\begin{equation} \label{yint}
\left. \log z \right|_{u}^{v} \;=\; \int_{u}^{v} y\;.
\end{equation}
Using~(\ref{5g}) in~(\ref{5ydef}) gives separate equations for the
amplitude and phase of $z$,
\[ \rho' \;=\; \rho\: {\mbox{Re}}\, y \;,\spc
\varphi' \;=\; {\mbox{Im}}\, y \:, \]
and integration gives
\begin{eqnarray}
\left. \log \rho \right|_{u}^{v} &=& \int_{u}^{v} {\mbox{Re}}\; y
\label{5A} \\
\left. \varphi \right|_{u}^{v} &=& \int_{u}^{v} {\mbox{Im}}\; y\;.
\label{5B}
\end{eqnarray}
Furthermore, the Wronskian~(\ref{wronski}) gives a simple algebraic
relation between $\rho$ and $y$. Namely, $w$ can be expressed by
$w = -{\mbox{Im}}\: (\overline{z}\: z' )
= \rho^2\: {\mbox{Im}}\:y$ and thus
\begin{equation}
\rho^2 \;=\; -\frac{w}{\mbox{Im}\:y}\;. \label{5C}
\end{equation}
Since~$\rho^2$ is positive and~$w$ is negative, we see that
\begin{equation}
{\mbox{Im}}\: y(u) > 0 \spc {\mbox{for all $u$}}. \label{5gg}
\end{equation}

\subsection{Invariant Disk Estimates}
\label{sec4}
We now explain a method for getting estimates for the complex
Riccati equation. This method was first used in~\cite{FS} for
estimates in the case where the potential is negative
(Lemma~\ref{lemmainv1}). Here we extend the method to the
situation when the potential is positive (Lemma~\ref{lemmainv2}).
For sake of clarity, we develop the method again from the beginning,
but we point out that the proof of Lemma~\ref{lemmainv1} is taken
from~\cite{FS}.
Let~$y(u)$ be a solution of the complex Riccati equation~(\ref{5c}).
We want to estimate the Euclidean distance of~$y$ to a given curve
$m(u)=\alpha + i \beta$ in the complex plane. A direct calculation
using~(\ref{5c}) gives
\begin{eqnarray*}
\lefteqn{ \frac{1}{2}\: \frac{d}{du}|y-m|^2 \;=\;
({\mbox{Re}}\, y - \alpha)\: ({\mbox{Re}}\, y - \alpha)'
+ ({\mbox{Im}}\, y - \beta)\: ({\mbox{Im}}\, y - \beta)' } \\
&=& ({\mbox{Re}}\, y - \alpha) \left[V  - ({\mbox{Re}}\, y)^2
+ ({\mbox{Im}}\, y)^2 - \alpha' \right]
- ({\mbox{Im}}\, y - \beta) \left[ 2\: {\mbox{Re}}\, y \:
{\mbox{Im}}\, y + \beta' \right] \\
&=& ({\mbox{Re}}\, y - \alpha) \left[ V  - ({\mbox{Re}}\, y)^2
- ({\mbox{Im}}\, y)^2 +2 \beta\:{\mbox{Im}}\, y - \alpha' \right]
\:+\:({\mbox{Re}}\, y - \alpha) \:2 ({\mbox{Im}}\, y - \beta)
\:{\mbox{Im}}\, y \\
&& - ({\mbox{Im}}\, y - \beta)
\left[\beta' + 2 \alpha\: {\mbox{Im}}\, y \right]
\:-\:({\mbox{Im}}\, y - \beta)\:2 ({\mbox{Re}}\, y - \alpha)
\:{\mbox{Im}}\, y \\
&=& ({\mbox{Re}}\, y - \alpha) \left[ V  - ({\mbox{Re}}\, y - \alpha)^2
- ({\mbox{Im}}\, y - \beta)^2 - \alpha^2+\beta^2-\alpha' \right] \\
&& - ({\mbox{Im}}\, y - \beta) \left[\beta' + 2 \alpha \beta \right]
- 2 \alpha \left( ({\mbox{Re}}\, y - \alpha)^2 + ({\mbox{Im}}\, y - \beta)^2
\right) .
\end{eqnarray*}
Choosing polar coordinates centered at~$m$,
\[ y \;=\; m + R e^{i \varphi} \:,\spc R \;:=\; |y-m| \:, \]
we obtain the following differential equation for~$R$,
\begin{equation} \label{Req}
R' + 2 \alpha R \;=\; \cos \varphi \left[
V -R^2 -\alpha^2 + \beta^2 - \alpha' \right]
- \sin \varphi \left[ \beta' + 2 \alpha \beta \right] .
\end{equation}

In order to use this equation for estimates, we assume
that~$\alpha$ is a given function (to be determined later).
With the abbreviations
\begin{equation} \label{Udef}
U \;=\; V - \alpha^2 - \alpha' \spc {\mbox{and}} \spc
\sigma(u) \;=\; \exp \left( 2 \int_0^u \alpha \right) ,
\end{equation}
the ODE~(\ref{Req}) can then be written as
\[ (\sigma R)' \;=\; \sigma \left[
U -R^2 + \beta^2 \right] \cos \varphi
- (\sigma \beta)'\:\sin \varphi \:. \]
To further simplify the equation, we want to arrange that the
square bracket vanishes. If~$U$ is negative, this can be achieved by
the ansatz
\begin{equation} \label{an1}
\beta \;=\; \frac{\sqrt{|U|}}{2} \left(T + \frac{1}{T} \right) , \spc
R \;=\; \frac{\sqrt{|U|}}{2} \left(T - \frac{1}{T} \right)
\spc (U<0),
\end{equation}
with~$T>1$ a free function. In the case~$U>0$, we make
similarly the ansatz
\begin{equation} \label{an2}
\beta \;=\; \frac{\sqrt{U}}{2} \left(T - \frac{1}{T} \right) , \spc
R \;=\; \frac{\sqrt{U}}{2} \left(T + \frac{1}{T} \right)
\spc (U>0)
\end{equation}
with a function~$T>0$. Using (\ref{an1}, \ref{an2}),
the ODE (\ref{Req}) reduces to the simple equation
$(\sigma R)'=-(\sigma \beta)' \sin \varphi$.
If we now replace this equation by a strict inequality,
\begin{equation} \label{din}
(\sigma R)' \;>\; -(\sigma \beta)' \sin \varphi \:,
\end{equation}
with~$R$ a general positive function,
the inequality~$|y-m| \leq R$ will be preserved as~$u$ increases.
In other words, the disk~$\overline{B_R}(m)$ will be an invariant region
for the flow of~$y$.  In the next two lemmas we specify the
function~$T$ in the cases~$U<0$ and~$U>0$, respectively.
To avoid confusion, we note that it is only a matter of convenience
to state the lemmas on the interval~$[0,\umax]$; by translation we
can later immediately apply the lemmas on any closed interval.

\begin{figure}[tbp]
\begin{center}
\input{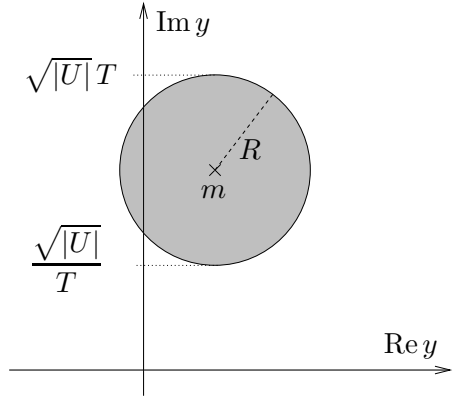}
\caption{Invariant disk estimate for~$U<0$.}
\label{fig1}
\end{center}
\end{figure}

\begin{Lemma} \label{lemmainv1}
Let~$\alpha$ be a real function on $[0,\umax]$ which is continuous and
piecewise $C^1$, such that the corresponding function~$U$, (\ref{Udef}),
is negative,
\[ U \leq 0 \quad {\mbox{on}} \quad [0, \umax]\: . \]
For a constant $T_0 \geq 1$ we introduce
the function~$T$ by
\begin{equation}
T(u) \;=\; T_0\: \exp \left( \frac{1}{2}\: {\mbox{TV}}_{[0,u)}
\log |\sigma^2 U| \right) \:, \label{1i}
\end{equation}
define the functions~$\beta$ and~$R$ by~(\ref{an1}) and
set~$m=\alpha + i \beta$.
If a solution $y$ of the complex Riccati equation~(\ref{5c}) satisfies
at~$u=0$ the condition
\begin{equation} \label{invest}
|y-m| \;\leq\; R \:,
\end{equation}
then this condition holds for all $u \in [0, \umax]$
(for illustration see Figure~\ref{fig1}).
\end{Lemma}
\Proof For $\varepsilon>0$ we set
\beq \label{Teps}
T_\varepsilon(u) \;=\; T_0\: \exp \left( \frac{1}{2}
\int_0^u \left| \frac{|\sigma^2 U|'}{|\sigma^2 U|} \right|
+ \varepsilon (1-e^{-u}) \right)
\eeq
and denote corresponding functions~$\alpha$, $R$, $m$,
and~$\sigma$ by an additional
subscript~$\varepsilon$. Since $T_\varepsilon(0)=T(0)$ and
$\lim_{\varepsilon \searrow 0} T_\varepsilon = T$, it suffices to show
that for all $\varepsilon>0$ the following statement holds,
\[ |y-m_\varepsilon|(0) \leq R_\varepsilon(0) \quad \Longrightarrow \quad
|y-m_\varepsilon|(u) \leq R_\varepsilon(u)\;
{\mbox{ for all $u \in [0, \umax]$}}. \]
In differential form, we get the sufficient condition
\[ |y-m_\varepsilon|(u) = R_\varepsilon(u)
\quad \Longrightarrow \quad
|y-m_\varepsilon|'(u) < R_\varepsilon'(u)\:. \]
According to~(\ref{din}), this last condition will be satisfied if
\beq \label{1m}
(\sigma_\varepsilon R_\varepsilon)' \;>\; |(\sigma_\varepsilon \beta_\varepsilon)'|\:.
\eeq
>>From now on we omit the subscripts~$\varepsilon$.

In order to prove~(\ref{1m}), we first use~(\ref{an1}, \ref{Udef})
to rewrite the functions $\sigma \beta$ and $\sigma R$ as
\beq \label{1l}
\left. \begin{array}{rcl}
\sigma \beta &=& \displaystyle \frac{1}{2} \left( \sqrt{|\sigma^2 U|}\:T +
\sqrt{|\sigma^2 U|}\:T^{-1} \right) \\[1em]
\sigma R &=& \displaystyle \frac{1}{2} \left( \sqrt{|\sigma^2 U|}\:T -
\sqrt{|\sigma^2 U|}\:T^{-1} \right) . \end{array} \quad \right\}
\eeq
By definition of $T_\varepsilon$~(\ref{Teps}),
\[ \frac{T'}{T} \;=\;
\frac{1}{2} \left| \frac{|\sigma^2 U|'}{|\sigma^2 U|} \right|
+ \varepsilon e^{-u} \:. \]
It follows that
\[ \left\{ \begin{array}{rclcl}
(\sqrt{|\sigma^2 U|}\: T^{-1})' &=&
-\varepsilon e^{-u} \: (\sqrt{|\sigma^2 U|}\: T^{-1})
&\quad& {\mbox{if $|\sigma^2 U|' \geq 0$}} \\[.5em]
(\sqrt{|\sigma^2 U|}\: T)' &=& \varepsilon e^{-u} \: (\sqrt{|\sigma^2 U|}\: T)
&\quad& {\mbox{if $|\sigma^2 U|' < 0$}}\:.
\end{array} \right. \]
Hence when we differentiate through~(\ref{1l}) and set
$\varepsilon=0$, either the first
or the second summand drops out in each equation, and we
obtain $(\sigma R)'=|\sigma \beta|'$. If $\varepsilon>0$, an inspection
of the signs of the additional terms gives~(\ref{1m}).
\QED

\begin{figure}[tbp]
\begin{center}
\input{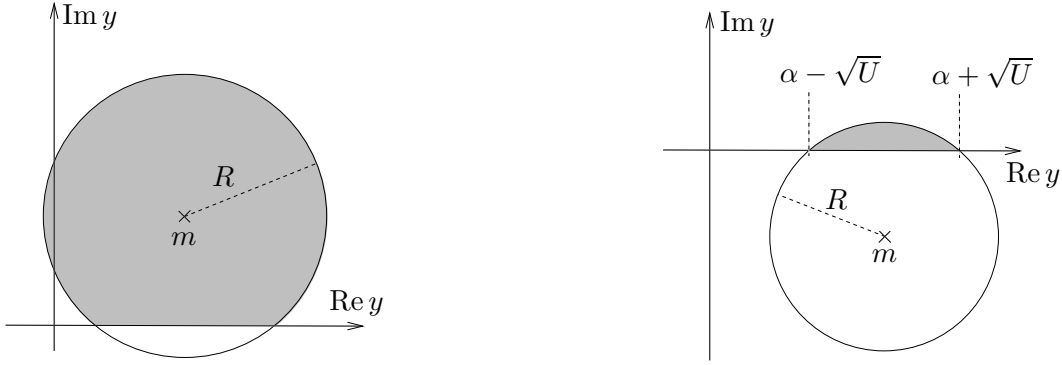}
\caption{Invariant disk estimate for~$U>0$,
in the cases $T>1$ (left) and $T<1$ (right).}
\label{fig2}
\end{center}
\end{figure}

\begin{Lemma} \label{lemmainv2}
Let~$\alpha$ be a real function on $[0,\umax]$ which is continuous and
piecewise $C^1$, such that the corresponding function~$U$, (\ref{Udef}),
satisfies on~$[0, \umax]$ the conditions
\beq \label{pcond}
U \geq 0 \qquad {\mbox{and}} \qquad
U' + 4 U \alpha \;\geq\; 0\:.
\eeq
For a constant $T_0 \geq 0$ we introduce
the function~$T$ by
\begin{equation}
T(u) \;=\; T_0\:\sqrt{\frac{U(0)}{\sigma^2 U}} \:, \label{1i2}
\end{equation}
define the functions~$\beta$ and~$R$ by~(\ref{an2}) and
set~$m=\alpha + i \beta$. If a solution $y$ of the complex Riccati equation~(\ref{5c}) satisfies
at~$u=0$ the condition
\[ |y-m| \;\leq\; R \:, \]
then this condition holds for all~$u \in [0, \umax]$
(see Figure~\ref{fig2}). Furthermore,
\beq \label{rel}
{\mbox{\rm{Re}}}\, y \;\geq\; \alpha - \sqrt{U} - T_0\: \frac{\sqrt{U(0)}}{2 \sigma}\:.
\eeq
\end{Lemma}
\Proof For~$\varepsilon>0$ we set
\[ T_\varepsilon \;=\; T_0 \:(\sigma^2 U)^{-\frac{1}{2}} \: (1-\varepsilon e^{-u}) \:. \]
Using~(\ref{an2}, \ref{Udef}) we can write the functions~$\sigma \beta$
and~$\sigma R$ as
\[ \left. \begin{array}{rcl}
\sigma \beta &=& \displaystyle -\frac{1}{2} \:\Big( T_0^{-1}\: \sigma^2 U\:
(1-\varepsilon e^{-u})^{-1} - T_0\: (1-\varepsilon e^{-u}) \Big) \\[1em]
\sigma R &=& \displaystyle \frac{1}{2} \:\Big( T_0^{-1}\: \sigma^2 U\:
(1-\varepsilon e^{-u})^{-1} + T_0\: (1-\varepsilon e^{-u}) \Big) , \end{array} \quad \right\} \]
where we again omitted the subscript~$\varepsilon$. Differentiation gives
\beq \label{pin}
(\sigma R)' \;>\; - (\sigma \beta)' \;=\;
\frac{1}{2}\: T_0^{-1} \left(\sigma^2 U \:(1-\varepsilon e^{-u})^{-1} \right)'
- \frac{1}{2}\: T_0 \left(1-\varepsilon e^{-u} \right)'\:.
\eeq
According to the second inequality in~(\ref{pcond}), the
function~$\sigma^2 U$ is strictly increasing and thus the expression
on the right of~(\ref{pin}) is positive for sufficiently
small~$\varepsilon$. Hence~(\ref{1m}) is satisfied.
Letting~$\varepsilon \to 0$, we obtain that the circle~$\overline{B}_R(m)$
is invariant.

\begin{figure}[tbp]
\begin{center}
\input{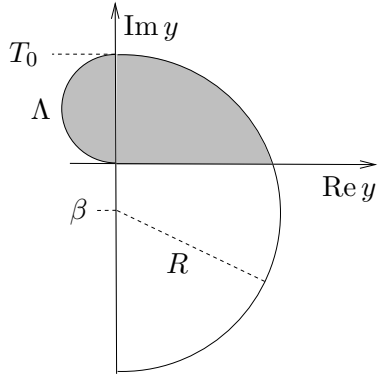}
\caption{Invariant region estimate for monotone~$V$.}
\label{fig4}
\end{center}
\end{figure}

In order to prove~(\ref{rel}) we note that in the case~$T<1$ the inequality
is obvious because even~${\mbox{Re}}\, y \leq \alpha - \sqrt{U}$ (see Figure~\ref{fig2}).
Thus we can assume~$T \geq 1$, and the estimate
\[ {\mbox{Re}}\, y \;\geq\; \alpha - R \;\geq\; \alpha - \frac{\sqrt{U}}{{2}}
\:( T + 2 ) \]
together with~(\ref{1i2}) gives the claim.
\QED
If the potential~$V$ is monotone increasing, by choosing~$\alpha \equiv 0$
we obtain the following simple estimate.
\begin{Corollary} \label{corsimple}
Assume that the potential~$V$ is monotone increasing on~$[0, \umax]$.
For a constant~$T_0>0$ with~$T_0^2 \geq -V(0)$ we introduce the functions
\beq \label{bTdef}
\beta \;=\; \frac{1}{2} \left( T_0 - \frac{V}{T_0} \right) , \spc
R \;=\; \frac{1}{2} \left( T_0 + \frac{V}{T_0} \right) .
\eeq
If a solution of the complex Riccati equation~(\ref{5c})
satisfies at~$u=0$ the condition
\[  y \;\in\; \Big\{ z \:|\: |z- i \beta| \leq R,\; {\mbox{\rm{Re}}}\, z,
{\mbox{\rm{Im}}}\, z \geq 0 \Big\} \cup
\left\{ z \:|\: \left| z - \frac{iT_0}{2} \right| \leq \frac{T_0}{2},\;
{\mbox{\rm{Re}}}\, z \leq 0 \right\} , \]
then this condition holds for all~$u \in [0, \umax]$ (see Figure~\ref{fig4}).
\end{Corollary}
{\Proof} Choosing~$\alpha \equiv 0$ and~$\beta$, $T$ according to~(\ref{bTdef}),
we know from Lemma~\ref{lemmainv1} and Lemma~\ref{lemmainv2} that the
circles~$|y-m| \leq R$ are invariant. Furthermore, we note that the arc~$\Lambda$
in Figure~\ref{fig4} is the flow line of the equation $y' + y^2 = 0$, and thus it
cannot be crossed from the right to the left when~$V$ is positive.
This gives the result in the case that~$V$ has no zeros. If~$V$ has a zero,
the invariant disks in the regions~$V \leq 0$ and~$V \geq 0$ coincide at
the zero of~$V$.
\QED

\begin{figure}[tb]
\begin{center}
\input{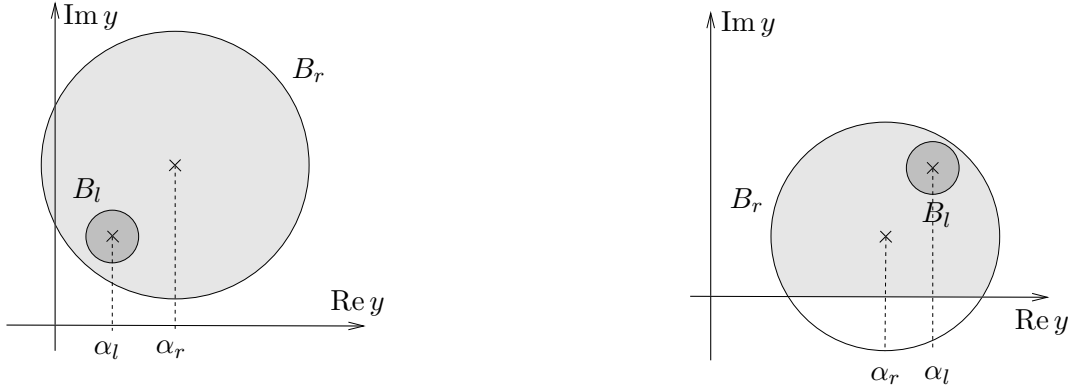}
\caption{Matching of invariant disks in the cases~$U_r<0$ (left)
and~$U_r>0$ (right).}
\label{fig3}
\end{center}
\end{figure}

The invariant disk estimates of Lemma~\ref{lemmainv1} and Lemma~\ref{lemmainv2}
can also be used if the functions~$\alpha$ and~$U$ have a discontinuity at some $v \in [0, \umax]$, i.e.
\[ \alpha_l \;:=\; \lim_{u \nearrow v} \alpha(u) \;\neq\;
\lim_{u \searrow v} \alpha(u) \;=:\; \alpha_r \:,\qquad
U_l \;:=\; \lim_{u \nearrow v} U(u) \;\neq\;
\lim_{u \searrow v} U(u) \;=:\; U_r \:. \]
In this case we choose the function~$T$ also to be discontinuous
at~$v$,
\[ T_l \;:=\; \lim_{u \nearrow v} T(u) \;\neq\;
\lim_{u \searrow v} T(u) \;=:\; T_r \:, \]
in such a way that the circle corresponding to~$(\alpha_r, U_r, T_r)$
contains that corresponding to~$(\alpha_l, U_l, T_l)$ (see Figure~\ref{fig3}).
In the next lemma we give sufficient ``jump conditions'' for this ``matching.''
\begin{Lemma} {\bf{(matching of invariant disks)}}
Suppose that~$U_l<0$. Depending on the sign of~$U_r$, we set
\begin{eqnarray}
T_r &=& T_l \;\frac{(\alpha_r-\alpha_l)^2 + |U_l + U_r|}
{\sqrt{|U_l|\: |U_r|}} \spc\spc\qquad\;\: {\mbox{if~$U_r < 0$}} \label{T1m} \\
T_r &=& T_l \;\frac{(\alpha_r-\alpha_l)^2 + |U_l + U_r| +
\sqrt{|U_l|\: U_r}}
{\sqrt{|U_l|\: U_r}} \spc {\mbox{if~$U_r>0$}} \label{T1p}
\end{eqnarray}
Let~$B_{l\!/\!r}$ be the disks with
centers~$m_{l\!/\!r} =\alpha_{l\!/\!r} + i \beta_{l\!/\!r}$
and radii~$R_{l\!/\!r}$ as
given by~(\ref{an1}) or~(\ref{an2}). Then~$B_l \subset B_r$.
\end{Lemma}
{\Proof} We must satisfy the condition~$R_r \geq |m_r-m_l| + R_l$.
Taking squares, we obtain the equivalent conditions~$R_r \geq R_l$ and
\[ (R_r - R_l)^2 \;\geq\; (\alpha_r-\alpha_l)^2 +
(\beta_r-\beta_l)^2 \:. \]
This last condition can also be written as
\beq \label{cm1}
(\alpha_r-\alpha_l)^2 + (\beta_l^2 - R_l^2) + (\beta_r^2 - R_r^2)
\;\leq\; 2 \:(\beta_l \beta_r - R_l R_r)\:.
\eeq

In the case~$U_r < 0$, we can substitute the ansatz~(\ref{an1})
into~(\ref{cm1}) to obtain the equivalent inequality
\[ (\alpha_r-\alpha_l)^2 + |U_l| + |U_r| \;\leq\;
\sqrt{|U_l| \: |U_r|} \left( \frac{T_r}{T_l} + \frac{T_l}{T_r} \right) . \]
Dropping the last summand on the right and solving for~$T_r$, we
obtain~(\ref{T1m}), which is thus a sufficient condition.

In the case~$U_r > 0$, we substitute~(\ref{an1}, \ref{an2})
into~(\ref{cm1}) to obtain the equivalent condition
\[ (\alpha_r-\alpha_l)^2 + |U_l| - U_r \;\leq\;
\sqrt{|U_l| \: U_r} \left( \frac{T_r}{T_l} - \frac{T_l}{T_r} \right) . \]
Using the inequality~$|U_l| - U_r \leq |U_l + U_r|$, replacing
the factor~$T_l/T_r$ on the right by one and solving for~$T_r$, we
obtain the sufficient condition~(\ref{T1p}).
\QED

\subsection{Bounds for the Wronskian and the Fundamental Solutions} \label{sec4.3}
We now consider the solutions~$\acute{\phi}$ and~$\grave{\phi}$ as defined in
Section~\ref{secholo} for~$\omega$ on the real axis and set
\[ \acute{y} \;=\; \frac{\acute{\phi}'}{\acute{\phi}}\:,\spc
\grave{y} \;=\; \frac{\grave{\phi}'}{\grave{\phi}} \:. \]
We keep~$k$ fixed. Since taking the complex conjugate of the separated
wave equation flips the sign of~$k$, we may assume that~$k \geq 0$. Then~$\omega_0$ as
defined by~(\ref{o0def}) is negative.
\begin{Prp} \label{prpode1}
If~$\omega \not \in [\omega_0, 0]$, the Wronskian
$w(\acute{\phi}, \grave{\phi})$ is non-zero.
\end{Prp}
{\Proof} According to~(\ref{Odef}), $\omega$ and~$\Omega$ have the same sign.
From~(\ref{5gg}) we know that the functions~$\acute{y}$ and~$\grave{y}$ both stay
either in the upper or lower half plane.
In view of the asymptotics~(\ref{abc1}, \ref{abc2}), we know that they must be
in be in opposite half planes. Thus
\[ w(\acute{\phi}, \grave{\phi}) \;=\; \acute{\phi} \:\grave{\phi}
\left(\acute{y} - \grave{y} \right) \;\neq\; 0 \:. \]

\vspace*{-.8cm}
\QED

In the case~$\omega \in (\omega_0, 0)$, we need the following global estimate
for large~$\lambda$.
\begin{Prp} \label{prpode2}
For any~$u_1 \in \R$ there are constants~$c, \lambda_0>0$ such that
\[ \left| \frac{\acute{\phi}(u)}{w(\acute{\phi}, \grave{\phi})} \right|
\;\leq\; \frac{c}{\lambda}
 \spc {\mbox{for all~$\lambda>\lambda_0, \:\omega \in (\omega_0, 0), \:u <u_1$}}\:. \]
\end{Prp}
The remainder of this section is devoted to the proof of this proposition.
Let~$u_1 \in \R$ and~$\omega \in (\omega_0, 0)$. Possibly by increasing~$u_1$
and~$\lambda_0$ we can clearly arrange that~$V$ is monotone decreasing on~$[u_1, \infty)$.
Then we have the following estimate.
\begin{Lemma} \label{lemmaeasy}
The functions~$\grave{\phi}$ and~$\grave{y}$ satisfy the
inequalities
\[ |\grave{\phi}(u)| \;\geq\; 1 \:,\quad {\mbox{\rm{Re}}}\, \grave{y}(u) \;\leq\; |\omega|
\spc {\mbox{on~$[u_1, \infty)$}}\:. \]
\end{Lemma}
{\Proof} From the asymptotics~(\ref{abc2}) we know that~$\lim_{u \to \infty}
\grave{y}(u) = -i \omega$. Thus for~$v$
sufficiently large, $|\grave{y}(v) - i |\omega|| < \varepsilon$, and we can apply Corollary~\ref{corsimple} on the
interval~$[u_1, v]$ backwards in~$u$ with~$T_0 = |\omega|+2 \varepsilon$.
Since~$\varepsilon$ can be chosen arbitrarily small, we conclude that
Corollary~\ref{corsimple} applies even on~$[u_1, \infty)$ with~$T_0=|\omega|$.
This means that
\[ 0 \;\leq\; {\mbox{Im}}\, \grave{y} \;\leq\; |\omega| \:,\quad
{\mbox{Re}}\, \grave{y} \;\leq\; |\omega| \spc {\mbox{on~$[u_1, \infty)$}}\:. \]
Finally, we use~(\ref{5C}) with~$w=i |\omega|$.
\QED

We now come to the estimates for~$\acute{\phi}$, which are more difficult because we need a
stronger result. The next lemma specifies the behavior of the potential
on~$(-\infty, 2 u_1]$.
\begin{Lemma} \label{lemma4.8}
For any~$u_1 \in \R$ there are constants~$c, \lambda_0$ such that
the potential~$V$ has for all~$\omega \in (\omega_0, 0)$ and all~$\lambda>\lambda_0$
the following properties. There are unique points~$u_- < u_0 < u_+ < u_1$ such that
\[ V(u_-) \;=\; -\frac{\Omega^2}{2} \:,\qquad V(u_0) \;=\; 0 \:,\qquad
V(u_+) \;=\; \Omega^2\:. \]
$V$ is monotone increasing on~$(-\infty, u_+]$.
Furthermore,
\begin{eqnarray}
u_+ - u_- &\leq& c \label{casy1} \\
\gamma \,u_+ &\geq& \log \Omega^2 - \log \lambda - c \label{casy3} \\
|V'|^{\frac{2}{3}} + |V''|^\frac{1}{2} &\leq& \frac{1}{4}\:|V| \spc {\mbox{on~$[u_+, 2 u_1]$}} ,
\label{casy2}
\end{eqnarray}
with~$\gamma$ as in~(\ref{Wbound}).
\end{Lemma}
{\Proof} We expand~$V$ in a Taylor series around the event horizon,
\[ V \;=\; -\Omega^2 + (\lambda+c_0) (r-r_1) + \lambda \:{\mathcal{O}}((r-r_1)^2)\:. \]
Hence for sufficiently large~$\lambda_0$ there are near the event horizon unique
points~$u_-$, $u_0$, $u_+$ where the potential has the required value.
Integrating~(\ref{51a}) we get near the event horizon the asymptotic formula
\[ u \;\sim\; \frac{1}{\gamma}\: \log(r-r_1) \:. \]
Getting asymptotic expansions for~$u_\pm$ we immediately obtain~(\ref{casy1}, \ref{casy3}).
Furthermore, using~(\ref{51a}) to transform $r$-derivatives into~$u$-derivatives,
we obtain in the region~$(r_1, r_1+\varepsilon) \cap (u_+, \infty)$ the estimates
\begin{eqnarray*}
\frac{\lambda}{c}\:e^{\gamma u} \;\leq\; V(u) &\leq& \lambda\,c\: e^{\gamma u} \\
|V'(u)|+|V''(u)| &\leq& \lambda\,c\: e^{\gamma u} \:,
\end{eqnarray*}
uniformly in~$\lambda$ and~$\omega$. Hence for sufficiently large~$\lambda_0$,
(\ref{casy2}) will be satisified near the event horizon.

In the region~$r>r_1+\varepsilon$ away from the event horizon, V is strictly
positive, $V > \lambda/c$, and since the derivatives of~$V$ can clearly be bounded
by~$|V'|+|V''| < c \lambda$, it follows that~(\ref{casy2}) is again satisfied.
\QED

First we apply Corollary~\ref{corsimple} on the interval~$(-\infty, u_-)$ to
obtain the following result.
\begin{Corollary} \label{cor1}
There is a constant~$c>0$ such that for all~$\omega \in (\omega_0, 0)$ and~$\lambda>\lambda_0$,
\[ \frac{\Omega}{2} \;\leq\; {\mbox{\rm{Im}}}\, y \;\leq\; \Omega \:,\quad
|{\mbox{\rm{Re}}}\, y| \;\leq\; \frac{\Omega}{2} \spc
{\mbox{on~$(-\infty, u_-]$}}\:. \]
\end{Corollary}
Also, at~$u=u_-$ we have an invariant disk with
\beq \label{id1}
\alpha_l\;=\; 0 \:,\qquad U_l\;=\; -\frac{\Omega^2}{2}\:, \qquad T_l \;=\; \sqrt{2}\:.
\eeq

On the interval~$[u_-, u_+]$ we use the method described in the next lemma.
\begin{Lemma} \label{lemmay3}
Assume that the potential~$V$ is monotone increasing on $[0, \umax]$.
We set
\[ \alpha \;=\; \sqrt{\max(2\:V(\umax), 0)} \]
and introduce for a given constant~$T_0>1$ the
functions~$U$, $\sigma$, $\beta$, $R$, and~$T$ by~(\ref{Udef}, \ref{an2}) and
\beq \label{Tsim}
T(u) \;=\; T_0  \:e^{2\alpha u} \:\frac{\sqrt{|U(0)|}}{\sqrt{|U(u)|}}\:.
\eeq
If a solution $y$ of the complex Riccati equation~(\ref{5c}) satisfies
at~$u=0$ the condition
\[ |y-m| \;\leq\; R \:, \]
then this condition holds for all~$u \in [0, \umax]$.
\end{Lemma}
{\Proof} By definition of~$\alpha$, the function~$U=V-\alpha^2$ is
negative and monotone increasing. Using furthermore
that~$\sigma=e^{2 \alpha u}$, we can estimate the total variation
in~(\ref{1i}) as follows,
\[ {\mbox{TV}}_{[0,u)} \log |\sigma^2 U| \;=\;
\int_0^u \left( 4 \alpha - \frac{|U|'}{|U|} \right)
\;=\; 4 \alpha u + \log |U(0)| - \log |U(u)| . \]
This gives~(\ref{Tsim}).
\QED
Thus we match the invariant disk~(\ref{id1}) to a disk with~$U_r=V(u_-)-\alpha_r^2$
and~$\alpha_r = \sqrt{2}\, \Omega$. From~(\ref{casy1}) we see that~$(u_+ - u_-)\, \alpha$
is uniformly bounded, and thus we obtain the following estimate.
\begin{Corollary} \label{cor2}
There is a constant~$c>0$ such that for all~$\omega \in (\omega_0, 0)$
and~$\lambda>\lambda_0$,
\[ \frac{\Omega}{c} \;\leq\; {\mbox{\rm{Im}}}\, y \;\leq\; c\,\Omega \:,\quad
|{\mbox{\rm{Re}}}\, y| \;\leq\; c\,\Omega \spc
{\mbox{on~$[u_-, u_+]$}}\:. \]
\end{Corollary}
At~$u=u_+$ we get an invariant disk with
\beq \label{id2}
0 \;\leq\; \alpha_l \;\leq\; \Omega\:,\quad
-c \, U_l \;=\; -\Omega^2 \:,\qquad
T_l \:\leq\; c\:.
\eeq

In the remaining interval~$[u_+, 2 u_1]$ an approximate solution
of the Schr{\"o}dinger equation~(\ref{5ode}) is available
from semi-classical analysis: the WKB wave function
\[ \phi(u) \;=\; V^{-\frac{1}{4}}\: \exp \left( \int^u \sqrt{V} \right) . \]
The corresponding function~$y$ is given by
\[ y(u) \;=\; \sqrt{V} - \frac{V'}{4V}\:. \]
In order to get an invariant disk estimate which quantifies the
exponential increase of~$\varphi$, we choose~$\alpha$ such that it also
becomes large as~$V \gg 0$. For technical simplicity, we choose
\beq \label{alphadef}
\alpha(u) \;=\; \frac{7}{8}\: \sqrt{V(u)}\:,
\eeq
giving rise to the following general result.
\begin{Lemma} \label{lemmay2}
Assume that the potential~$V$ is positive on~$[0, \umax]$ and that
\beq \label{y2assum}
|V'(u)| \;\leq\; \frac{1}{2}\: V(u)^{\frac{3}{2}} \:,\spc
|V''(u)| \;\leq\; \frac{1}{4}\: V(u)^2\:.
\eeq
We introduce for a given constant~$T_0>0$ the
functions~$\alpha$, $U$, $\sigma$, $\beta$, $R$, and~$T$
by~(\ref{alphadef}, \ref{Udef}, \ref{an2}, \ref{1i2}).
If a solution $y$ of the complex Riccati equation~(\ref{5c}) satisfies
at~$u=0$ the condition
\[ |y-m| \;\leq\; R \:, \]
then this condition holds for all~$u \in [0, \umax]$. Furthermore,
\beq \label{rel2}
{\mbox{\rm{Re}}}\, y \;\geq\; \frac{\sqrt{V}}{8} \:-\: \frac{T_0}{2}\,|\Omega|\:.
\eeq
\end{Lemma}
{\Proof} A short calculation yields
\begin{eqnarray*}
U &=& V - \alpha^2 - \alpha' \;=\; \frac{15}{64}\: V
- \frac{7}{16}\: \frac{V'}{\sqrt{V}} \\
U' + 4 \alpha U &=&
\frac{105}{128}\: V^{\frac{3}{2}} - \frac{83}{64}\: V'
+ \frac{7}{32}\: \frac{V'^2}{V^{\frac{3}{2}}}
- \frac{7}{16}\: \frac{V''}{\sqrt{V}} \;.
\end{eqnarray*}
Using~(\ref{y2assum}) we obtain the estimates
\[ \frac{V}{64} \;\leq\; U \;\leq\; \frac{V}{2}
\spc{\mbox{and}}\spc U' + 4 \alpha U \;\geq\; \frac{V^{\frac{3}{2}}}{16}\:. \]
Hence the conditions~(\ref{pcond}) are satisified, and
Lemma~\ref{lemmainv2} applies. The inequality~(\ref{rel2}) follows from~(\ref{rel}),
the just-derived upper bound for~$U$ and the fact that~$\sigma \geq 1$.
\QED

Matching the invariant disk~(\ref{id2}) to the invariant disk
with~$\alpha_r=\alpha(u_r)$ and~$U_r=V(u_r)-\alpha^2(u_r)-\alpha'(u_r)$
with~$\alpha$ according to~(\ref{alphadef}), we obtain
\beq \label{Trrel}
U_r \;\leq\; \Omega^2 \:,\spc T_r \;\leq\; c \:.
\eeq
We can then apply the last lemma on the interval~$[u_+, 2u_1]$. \\[1em]
{\em{Proof of Proposition~\ref{prpode2}. }}
Suppose that~$u < u_1$. Using the
definition of~$\acute{y}$ and~$\grave{y}$, we can rewrite the Wronskian as
\[ w(\acute{\phi}, \grave{\phi}) \;=\;
\acute{\phi}\: \grave{\phi}\:
(\acute{y} - \grave{y}) \:. \]
Applying Lemma~\ref{lemmaeasy} at~$u=2u_1$ gives
\beq \label{inter1}
\left| \frac{\acute{\phi}(u)}{w(\acute{\phi}, \grave{\phi})} \right|
\;\leq\; \left| \frac{\acute{\phi}(u)}{\acute{\phi}(2 u_1)} \right|
\: \frac{1}{{\mbox{Re}}\,\acute{y}(2 u_1) - |\omega|} \:.
\eeq

We combine~(\ref{rel2}) with $T_0=T_r$ satisfying~(\ref{Trrel}) to get
\beq \label{corin0}
{\mbox{\rm{Re}}}\, \acute{y} \;\geq\; \frac{\sqrt{V}}{8} \:-\: c \Omega \:.
\eeq
Since the potential~$V$ is strictly positive on the interval~$[u_1, 2u_1]$,
we can, possibly by increasing~$\lambda_0$ and~$c$, arrange that
\beq \label{Vsqrt}
\sqrt{V} \;\geq\; \frac{\sqrt{\lambda}}{c} \spc {\mbox{on~$[u_1, 2 u_1]$}}
\eeq
and thus also that
\beq
{\mbox{\rm{Re}}}\, \acute{y} \;\geq\; \frac{\sqrt{\lambda}}{16\,c} \spc {\mbox{on~$[u_1, 2 u_1]$}} \label{corin1} \:.
\eeq
This inequality allows us to bound the fraction in~(\ref{inter1}),
\beq \label{inter2}
\left| \frac{\acute{\phi}(u)}{w(\acute{\phi}, \grave{\phi})} \right|
\;\leq\; \left|\frac{\acute{\phi}(u)}{\acute{\phi}(2 u_1)} \right| \:.
\eeq

Thus it remains to control the last quotient. We omit the accent and
use the notation~$\rho=|\phi|$.
In the case~$u < u_+$, we can use~(\ref{5C}),
\[ \frac{\rho(u)^2}{\rho(u_+)^2} \;=\; \frac{{\mbox{Im}}\, y(u_+)}{{\mbox{Im}}\, y(u)} \:, \]
and the last quotient is controlled from above and below by Corollary~\ref{cor1}
and Corollary~\ref{cor2}. Hence, rewriting the quotient on the right of~(\ref{inter2}) as
\[ \frac{\rho(u)}{\rho(2 u_1)} \;=\; \frac{\rho(u)}{\rho(u_+)}\: \frac{\rho(u_+)}{\rho(2 u_1)}\:, \]
it remains to consider the case~$u \geq u_+$.
Applying~(\ref{5A}) and~(\ref{corin0}), we obtain
\[ A \;:=\; \log \left| \frac{\phi(u)}{\phi(2 u_1)} \right|
\;=\; -\int_u^{2 u_1} {\mbox{Re}}\: y(u)
\;\leq\;
c\, \Omega\:(2 u_1 - u_+) \:-\: \frac{1}{8} \int_u^{2 u_1} \sqrt{V} \:. \]
Now we use~(\ref{casy3}) and the fact that the function~$\Omega\, \log \Omega$ is
bounded,
\[ A \;\leq\; c\:\log \lambda \:-\: \frac{1}{8} \int_u^{2 u_1} \sqrt{V} \:. \]
Estimating the last summand with~(\ref{Vsqrt}),
\[ \int_u^{2 u_1} \sqrt{V} \;\geq\; \int_{u_1}^{2 u_1} \sqrt{V}
\;\geq\; \frac{\sqrt{\lambda}}{c} \:u_1\:, \]
we conclude that for large~$\lambda$ this summand dominates
the term~$c \log \lambda$, and thus~(\ref{inter2}) decays in~$\lambda$
even like~$\exp(-\sqrt{\lambda}/c)$.
\QED

\section{Contour Deformations to the Real Axis}
\setcounter{equation}{0}
In this section we fix the angular momentum number~$k$ throughout and omit
the angular variable~$\varphi$. We can again assume
without loss of generality that~$k \geq 0$.
Also, since here we are interested
in the situation only locally in~$u$, we evaluate weakly.
Thus we write the integral representation~(\ref{intrep})
for compactly supported initial data~$\Psi_0$ and a test
function~$\eta \in C^\infty_0(\R \times S^2)^2$ as
\beq \label{intk}
\bra \eta,\,\Psi(t)\ket \;=\; -\frac{1}{2\pi i} \sum_{n \in \sN}\;
\lim_{\varepsilon \searrow 0} \left( \int_{C_\varepsilon} -\int_{\overline{C_\varepsilon}}
\right) d\omega\; e^{-i \omega t} \:\bra \eta,
Q_{n}(\omega)\: S_{\infty}(\omega)\: \Psi_0 \ket \:.
\eeq
The integration contour in~(\ref{intk}) can be moved to the real
axis provided that the integrand is continuous. In the next lemma
we specify when this is the case and simplify the integrand.
For~$\omega$ real, the complex conjugates of~$\acute{\phi}$
and~$\grave{\phi}$ are again solutions of the ODE. Thus, apart from
the exceptional cases~$\omega \in \{0, \omega_0\}$, we can
express~$\grave{\phi}$ as a linear combination of~$\acute{\phi}$
and~$\overline{\acute{\phi}}$, \beq \label{trans} \grave{\phi} \;=\;
\alpha\: \acute{\phi} + \beta\: \overline{\acute{\phi}} \spc (\omega
\in \R \setminus \{0, \omega_0\}) \:. \eeq The complex
coefficients~$\alpha$ and~$\beta$ are called {\em{transmission
coefficients}}. The Wronskian of~$\acute{\phi}$ and~$\grave{\phi}$
can then be expressed by
\beq \label{wroneq} w(\acute{\phi},
\grave{\phi}) \;=\; \beta \:w(\acute{\phi}, \overline{\acute{\phi}})
\;=\; 2 i \Omega\: \beta\:,
\eeq
where in the last step we used the asymptotics~(\ref{abc1}). Furthermore,
it is convenient to introduce the real fundamental solutions
\[ \phi_1 \;=\; {\mbox{\rm{Re}}}\, \acute{\phi} \:,\spc
\phi_2 \;=\; {\mbox{\rm{Im}}}\, \acute{\phi}\:, \]
and to denote the corresponding solutions of the wave equation in Hamiltonian
form by~$\Psi_{1\!/\!2}^{\omega n}$.

\begin{Lemma} \label{lemma5.1}
If the Wronskian~$w(\acute{\phi}, \grave{\phi})$ is
non-zero at~$\omega \in \R \setminus \{0, \omega_0\}$, then the integrand in~(\ref{intk}) is
continuous at~$\omega$ and
\beq \label{explicit}
\big( \lim_{\varepsilon \nearrow 0} - \lim_{\varepsilon \searrow 0} \big)
(Q_{n}(\omega+i \varepsilon)\: S_\infty(\omega+i \varepsilon) \:\Psi)(r, \vartheta)
\;=\; -\frac{i}{\omega \Omega} \sum_{a,b=1}^2 t^{\omega n}_{ab}\: \Psi_a^{\omega n}
\bra \Psi_b^{\omega n}, \Psi \ket \:,
\eeq
where the coefficients~$t_{ab}$ are given by
\beq \label{tabdef}
t_{11} \;=\; 1 + {\mbox{\rm{Re}}}\, \frac{\alpha}{\beta} \:,\qquad
t_{12} \;=\; t_{21} \;=\; -{\mbox{\rm{Im}}}\, \frac{\alpha}{\beta} \:,\qquad
t_{22} \;=\; 1 - {\mbox{\rm{Re}}}\, \frac{\alpha}{\beta}\:.
\eeq
\end{Lemma}
{\Proof} We start from the explicit formula for the operator product~$Q_{n} S_\infty$
given in~\cite[Proposition~5.4]{FKSY}. Since the angular operator~$Q_{n}(\omega+i \varepsilon)$
can be diagonalized for~$\varepsilon$ sufficiently small, the kernel~$g(u,u')$ is simply
the Green's function of the radial ODE, i.e. for~$\omega$ in the lower half plane,
\beq \label{Green}
g(u,u') \;:=\; \frac{1}{w(\acute{\phi}, \grave{\phi})} \:\times \left\{
\begin{array}{cl} \acute{\phi}(u)\: \grave{\phi}(u') & {\mbox{if $u \leq u'$}} \\
\grave{\phi}(u)\: \acute{\phi}(u') & {\mbox{if $u > u'$}}\:. \end{array} \right.\:,
\eeq
whereas the formula in the upper half plane is obtained by complex conjugation.
Using that~$\lim_{\varepsilon \nearrow 0} \acute{\phi} = \lim_{\varepsilon \searrow 0} \overline{\acute{\phi}}$
and~$\lim_{\varepsilon \nearrow 0} \grave{\phi} = \lim_{\varepsilon \searrow 0} \overline{\grave{\phi}}$,
we find that
\[ \big( \lim_{\varepsilon \nearrow 0} - \lim_{\varepsilon \searrow 0} \big) \:g(u,u')
\;=\; 2i\: {\mbox{Im}}\: g(u,u') \:, \]
and a short calculation using~(\ref{trans}, \ref{wroneq}) gives
\[ \big( \lim_{\varepsilon \nearrow 0} - \lim_{\varepsilon \searrow 0} \big) \:g(u,u')
\;=\; -\frac{i}{\Omega} \sum_{a,b=1}^2 t_{ab} \: \phi^a(u)\: \phi^b(u') \]
with~$t_{ab}$ according to~(\ref{tabdef}).

Except for the function~$g(u,u')$, all the functions appearing in the
formula for~$Q_n S_\infty$ in~\cite[Proposition~5.4]{FKSY} are continuous on the real axis.
A direct calculation shows that
\begin{eqnarray*}
\lefteqn{ \big( \lim_{\varepsilon \nearrow 0} - \lim_{\varepsilon \searrow 0} \big)
(Q_{k,n}(\omega+i \varepsilon)\: S_\infty(\omega+i \varepsilon) \:\Psi) } \\
&=&-\frac{i}{\omega \Omega} \sum_{a,b=1}^2 t_{ab}\:
\Psi_a \: \bra \Psi_b,\: \left( \!\!\begin{array}{cc}
(\omega - \beta) \omega & 0 \\ 0 & 1
\end{array} \!\!\right) \Psi \ket_{L^2(d\mu)}
\end{eqnarray*}
with~$d\mu$ given by~(\ref{mudef}).
Since the~$\Psi_b$ are eigenfunctions of the Hamiltonian,
we know according to~(\ref{HAb}) that~$A \Psi_b = (\omega-\beta)\omega \Psi_b$.
Using furthermore that the operator~$A$ is symmetric on~$L^2(d\mu)$,
we conclude that
\[ \bra \Psi_b,\: \left( \!\!\begin{array}{cc}
(\omega - \beta) \omega & 0 \\ 0 & 1
\end{array} \!\!\right) \Psi \ket_{L^2(d\mu)}
\;=\; \bra \Psi_b,\: \left( \!\!\begin{array}{cc}
A & 0 \\ 0 & 1
\end{array} \!\!\right) \Psi \ket_{L^2(d\mu)}
\;=\; \bra \Psi_b,\: \Psi \ket\:, \]
where in the last step we used~(\ref{ESP}).
\QED

\begin{figure}[tbp]
\begin{center}
\input{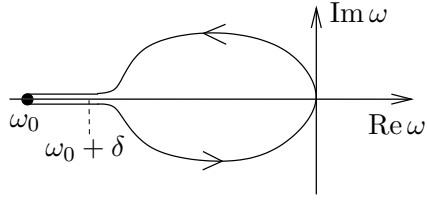}
\caption{The integration contour~$D_\delta$.}
\label{fig5}
\end{center}
\end{figure}

Let us now consider for which values of~$\omega$ and~$n$ the contour can be moved to
the real axis. According to Proposition~\ref{prpode1}, the Wronskian~$w(\acute{\phi},
\grave{\phi})$ is non-zero unless~$\omega \in [\omega_0, 0]$. We now analyze
carefully the exceptional cases~$\omega=0, \omega_0$. From Theorem~\ref{thmh1},
Theorem~\ref{thmh2} and Theorem~\ref{thmc1} we know that the functions~$\acute{\phi}$
and~$\phi_\omega = \omega^\mu \grave{\phi}$ are continuous for all~$\omega \in \R$.
If~$ak=0$ and~$\omega=0$, the functions~$\acute{\phi}$ and~$\phi_\omega$ degenerate to
real solutions with the asymptotics
\[ \lim_{u \to -\infty} \acute{\phi}(u) \;=\; 1\:,\spc
\lim_{u \to \infty} u^{\mu - \frac{1}{2}}\: \phi_0(u) \;=\; \frac{\Gamma(\mu)}{\sqrt{\pi}}\:. \]
Noting that the function
\[ \partial_u \sqrt{r^2+a^2} \;=\;
\frac{r\:\Delta}{(r^2+a^2)^{\frac{3}{2}}} \;=\; \frac{r}{\sqrt{r^2+a^2}}
\left(1 - \frac{2Mr}{r^2+a^2} \right) \]
is monotone increasing, the potential~$V$, (\ref{5V}), is everywhere positive.
Hence solutions of the Schr{\"o}dinger equation~(\ref{5ode}) are convex. This implies
that the functions~$\acute{\phi}$ and~$\grave{\phi}$ do not coincide, and
thus their Wronskian is non-zero. As a consequence, the Green's function~(\ref{Green}),
and thus the whole integrand in~(\ref{intk}), is bounded and continuous near~$\omega=0$
(note that~(\ref{Green}) is invariant under rescalings of~$\grave{\phi}$, and thus
we can in this formula replace~$\grave{\phi}$ by~$\phi_\omega$).
In the case~$ak \neq 0$ and~$\omega=0$, the function~$\phi_0$ is real,
whereas~$\acute{\phi}$ is complex, and thus~$w(\acute{\phi}, \phi_0) \neq 0$.
If on the other hand~$ak \neq 0$ and~$\omega=\omega_0$, $\acute{\phi}$ is real
and~$\grave{\phi}$ is complex, and again~$w(\acute{\phi}, \phi_0) \neq 0$.
Hence the integrand in~(\ref{intk}) is continuous and bounded at the points~$\omega=0,
\omega_0$. We conclude that for every~$n \in \N$, the integrand
in~(\ref{intk}) is continuous on an open neighborhood of
on~$\omega \in \R \setminus (\omega_0, 0)$. Furthermore, according to
Proposition~\ref{prpode2}, $w(\acute{\phi}, \grave{\phi}) \neq 0$ if~$\omega
\in (\omega_0, 0)$ and~$\lambda$ is sufficiently large. We have thus proved
the following result.
\begin{Prp} \label{prpcont}
There is~$\delta>0$ and~$n_0 \in \N$ such that for every~$\Psi_0 \in C^\infty_0(\R \times S^2)^2$, the completeness relation
\begin{eqnarray*}
\Psi_0 &=& \frac{1}{2 \pi}
\left( \sum_{n=0}^\infty \int_{\R \setminus [\omega_0, 0]} + \sum_{n>n_0} \int_{\omega_0}^0 \right)
\frac{d\omega}{\omega \Omega}
\sum_{a,b=1}^2 t^{\omega n}_{ab}\: \Psi_a^{\omega n} \: \bra \Psi_b^{\omega n}, \Psi_0 \ket \\
&&+ \sum_{n \leq n_0} \oint_{D_\delta}
(Q_{n}\: S_\infty\: \Psi_0) \:d\omega
\end{eqnarray*}
holds, with the contour~$D_\delta$ as in Figure~\ref{fig5}.
\end{Prp}
We point out that the contour~$D_\delta$ passes along the line
segment~$[\omega_0, \omega_0+\delta)$ twice, once as the limit of the
contour in the lower half plane, and once as limit of the contour in
the upper half plane. These two integrals can be combined to
one integral over~$[\omega_0, \omega_0+\delta)$ with the integrand
given by~(\ref{explicit}).

Let us now consider how the remaining contour integrals over~$C_\varepsilon$
can be moved to the real line.
According to Theorems~\ref{thmh1} and~\ref{thmh2}, the functions~$\acute{\phi}$
and~$\grave{\phi}$ have for every~$n \leq n_0$ and for
every~$\omega \in (\omega_0, 0)$ a holomorphic
extension to a neighborhood of~$\omega$. Thus their Wronskian is also
holomorphic in this neighborhood, and consequently they can have only isolated
zeros of finite order. Since~$w(\acute{\phi}, \grave{\phi}) \neq 0$
for~$\omega$ near~$0$ and~$\omega_0$, we conclude that the numbers of zeros
must be finite. Since we only need to consider a finite number of angular momentum modes, there is at most a finite number of points~$\omega_1, \ldots, \omega_K \in (\omega_0, 0)$, $K \geq 0$,
where any of the Wronskians~$w(\acute{\phi}_n, \grave{\phi}_n)$ has a zero. We denote the maximum of the orders of these zeros at~$\omega_i$ by~$l_i \in \N$.

The above zeros of the Wronskian lead to poles in the integrand of~(\ref{intk})
and correspond to radiant modes. We will prove in Section~\ref{secsuper}
by contradiction that these radiant modes are actually absent. Therefore, we
now make the assumption that there are radiant modes, i.e.\ that
the Wronskians $w(\acute{\phi}_n, \grave{\phi}_n)$ have at least one zero
on the real axis.
As a preparation for the analysis of Section~\ref{secsuper}, we now choose a special
configuration where radiant modes appear, but in the simplest possible way.
We choose new initial data
\beq \label{p0def}
\Phi_0 \;=\; {\mathcal{P}}(H)\: \Psi_0\:,
\eeq
where~${\mathcal{P}}$ is the polynomial
\[ {\mathcal{P}}(x) \;=\; \omega\, (\omega-\omega_0)\; (x-\omega_1)^{l_1-1} \prod_{i=2}^K
(x-\omega_i)^{l_i} \:. \]
Then~$\Phi_0$ again has compact support, and using the spectral calculus,
the corresponding solution~$\Phi(t)$ of the Cauchy
problem is obtained from~(\ref{intk}) by multiplying the integrand
by~${\mathcal{P}}(\omega)$. Then the poles of the integrand
at~$\omega_2, \ldots, \omega_K$ disappear, and at~$\omega_1$ a simple
pole remains. Subtracting this pole, the integrand becomes analytic,
whereas for the pole itself we get a contour integral which can be
computed with residues.

Let us summarize the result of the above construction with a compact notation.
For a test function~$\eta \in C^\infty_0(\R \times S^2)^2$ we introduce
the vectors~$\eta^{\omega n}$ by
\[ \eta^{\omega n} \;=\; (\eta^{\omega n}_1, \eta^{\omega n}_2)
\spc {\mbox{where}} \spc \eta^{\omega n}_a \;=\; \bra \Psi^{\omega n}_a, \eta \ket \:. \]
\begin{Prp} \label{prpsrm}
Assume that there are radiant modes, $K \geq 1$. Then the Cauchy development~$\Phi(t)$
of the initial data~(\ref{p0def}) satisfies the relation
\[ \bra \eta, \Phi(t) \ket \;=\; \frac{1}{2 \pi} \sum_{n \in \N}
\int_{-\infty}^\infty e^{-i \omega t}\: \lbra \eta^{\omega n}, T^{\omega n} \Phi^{\omega n}
\lket_{\C^2} \:d\omega
\;+\; e^{-i \omega_1 t} \sum_{n \leq n_0}\:
\lbra \eta^{\omega_1 n}, \sigma^n \lket_{\C^2} \:. \]
Here~$\omega_1 \in (\omega_0, 0)$. The~$(\sigma^n)_{n=1,\ldots, n_0}$ are
vectors in~$\C^2$, at least one of which
is non-zero. The matrices~$T^{\omega n}$ have the following properties,
\begin{description}
\item[(1)] If~$\omega \not \in [\omega_0, 0]$ or~$n > n_0$,
\beq \label{Tondef}
(T^{\omega n})_{ab} \;=\; t^{\omega n}_{ab}
\eeq
with~$t_{ab}$ according to~(\ref{tabdef}).
\item[(2)] For each~$n$, the function~$T^{\omega n}$ is continuous in~$\omega \in \R$
and analytic in~$(\omega_0, 0)$.
\end{description}
\end{Prp}

\section{Energy Splitting Estimates}
\setcounter{equation}{0}
In this section, we consider the family of test functions
\[ \eta_L(u) \;=\; \eta(u+L) \]
for a fixed~$\eta \in C^\infty_0(\R \times S^2)^2$. Our goal
is to control the inner product~$\bra \eta_L, \Phi(t) \ket$
in the limit~$L \to \infty$ when the support of~$\eta_L$
moves towards the event horizon.
Our method is to split up the inner product into
a positive and an indefinite part. Once the indefinite part is bounded
using the ODE estimates of Section~\ref{secode}, we can use the
Schwarz inequality and energy conservation to also control the positive part.

We choose~$u_1 \in \R$ and general test functions~$\eta, \zeta \in C^\infty_0((-\infty, u_1) \times
S^2)^2$ which are supported to the left of~$u_1$ (for later use we often work
more generally with~$\zeta$ instead of~$\Phi_0$).
Since for each fixed~$n$, the~$T^{\omega n}$ are continuous and the
eigensolutions~$\Psi^{\omega n}_a(u)$ are, according to Theorem~\ref{thmh1},
also continuous in~$\omega$, uniformly for~$u \in (-\infty, u_1)$, we have no
difficulty controlling the expressions~$\lbra \eta^{\omega n}, T^{\omega n}
\zeta^{\omega n} \lket_{\C^2}$ for~$n \leq n_0$ and~$\omega \in [\omega_0, 0]$.
Hence we only need to consider the case when the matrix~$T^{\omega n}$ is given
by~(\ref{Tondef}). Using~(\ref{tabdef}), the eigenvalues~$\lambda_\pm$
of this matrix are
\beq \label{ldef}
\lambda_\pm \;=\; 1 \pm \left| \frac{\alpha}{\beta} \right| .
\eeq
In order to determine the sign of these eigenvalues, we first use the
asymptotics~(\ref{abc1}, \ref{abc2}) to compute the Wronskians
$w(\grave{\phi}, \overline{\grave{\phi}}) = -2i \omega$
and~$w(\acute{\phi}, \overline{\acute{\phi}}) = 2i \Omega$.
Furthermore, we obtain from~(\ref{trans}) and its complex conjugate that
\[ w(\grave{\phi}, \overline{\grave{\phi}}) \;=\; (|\alpha|^2-|\beta|^2)\:
w(\acute{\phi}, \overline{\acute{\phi}}) \:. \]
Combining these identities, we find that
\beq
|\alpha|^2 - |\beta|^2 \;=\; -\frac{\omega}{\Omega} \label{abrel}
\eeq
From~(\ref{ldef}, \ref{abrel}) we see that in the case~$\omega \not \in [\omega_0, 0]$,
where~$\omega$ and~$\Omega$ have the same sign, the eigenvalues~$\lambda_\pm$
are both positive. However, if~$\omega \in (\omega_0, 0)$, one of the eigenvalues is
negative. This result is not surprising, because the lack of positivity corresponds
to the fact that for~$\omega \in [\omega_0, 0]$ the energy density can be negative
inside the ergosphere. In the case when~$T^{\omega n}$ is not positive, we decompose
it into the difference of two positive matrices,
\[ T^{\omega n} \;=\; T^{\omega n}_+ - T^{\omega n}_- \spc
{\mbox{for~$\omega \in (\omega_0, 0),\: n >n_0$}}\:, \]
where
\[ T^{\omega n}_- \;=\; -\lambda_-\:\1\:. \]

In the next lemma we bound the integral over~$T^{\omega n}_-$ using ODE techniques.
\begin{Lemma} \label{lemmaneg}
For any~$\varepsilon>0$ we can, possibly by increasing~$n_0$,
arrange that for all~$L \leq 0$,
\[ \sum_{n > n_0} \int_{\omega_0}^0
\left| \lbra \eta^{\omega n}_L, T^{\omega n}_- \zeta^{\omega n} \lket_{\C^2}
\right| \:d\omega\;\leq\; \varepsilon\:. \]
\end{Lemma}
{\Proof} Using~(\ref{abrel}, \ref{wroneq}) we can estimate the
norm of~$T_-$ by
\[  \|T_-\| \;=\; |\lambda_-| \;=\;
\frac{|\alpha|^2-|\beta|^2}{|\beta|\: (|\alpha|+|\beta|)}
\;\leq\; \left|\frac{\omega}{\Omega}\right| \frac{1}{2\:|\beta|^2}
\;=\; \frac{2 |\omega \Omega|}{|w(\acute{\phi}, \grave{\phi})|^2}\:. \]
Hence
\[ \sum_{n > n_0} \int_{\omega_0}^0
\left| \lbra \eta^{\omega n}_L, T^{\omega n}_- \zeta^{\omega n} \lket_{\C^2}
\right| \:d\omega \;\leq\; 2 \sum_{n > n_0} \int_{\omega_0}^0
\frac{|\eta^{\omega n}_L|}{|w(\acute{\phi}, \grave{\phi})|}\:
\frac{|\zeta^{\omega n}|}{|w(\acute{\phi}, \grave{\phi})|}\:
|\omega \Omega|\: d\omega\:. \]
Writing out the energy scalar product
using~\cite[eq.~(2.14)]{FKSY}
and expressing the fundamental
solutions~$\Psi^{\omega n}_a$ in terms of the radial solution~$\acute{\phi}$, one sees that
\[ |\eta_L^{\omega n}| \;\leq\; c\: \sup_{\R} |\eta_L \:\acute{\phi}|\:,\spc
|\zeta^{\omega n}| \;\leq\; c\: \sup_{\R} |\zeta \:\acute{\phi}|\:, \]
where the constant~$c=c(\omega)$ is independent of~$\lambda$.
Now we apply Proposition~\ref{prpode2} and use that the eigenvalues~$\lambda_n$
grow quadratically in~$n$, (\ref{quadra}).
\QED

\begin{Lemma} \label{lemmale}
There is a constant~$C>0$ such that for all~$L \geq 0$,
\[ \sum_{n \in \N} \int_{\R \setminus [\omega_0, 0]}
\left| \lbra \eta^{\omega n}_L, T^{\omega n} \zeta^{\omega n} \lket_{\C^2}
\right| \, d\omega \;\leq\; C\:. \]
\end{Lemma}
{\Proof}
First of all, using the positivity of the matrix~$T$,
\begin{eqnarray*}
\lefteqn{  \sum_{n \in \N} \int_{\R \setminus [\omega_0, 0]}
\left| \lbra \eta^{\omega n}_L, T^{\omega n} \zeta^{\omega n} \lket_{\C^2} \right| \,
d\omega } \\
&\leq& \frac{1}{2} \sum_{n \in \N} \int_{\R \setminus [\omega_0, 0]}
\left( \lbra \eta^{\omega n}_L, T^{\omega n} \eta^{\omega n}_L \lket_{\C^2}
\:+\: \lbra \zeta^{\omega n}, T^{\omega n} \zeta^{\omega n} \lket_{\C^2} \right)
\, d\omega .
\end{eqnarray*}
The two summands can be treated in exactly the same way; we treat the summand
involving~$\eta^{\omega n}_L$ because of the additional $L$-dependence.
Applying Proposition~\ref{prpcont} and dropping all negative terms, we get
\begin{eqnarray*}
\lefteqn{ \int_{\R \setminus [\omega_0, 0]} \lbra \eta^{\omega n}_L,
T^{\omega n}_+ \eta^{\omega n}_L \lket_{\C^2} \: d\omega
\;\leq\; \bra \eta_L, H\,(H-\omega_0)\, \eta_L \ket } \\
&&+ \sum_{n > n_0} \int_{\omega_0}^0 \lbra \eta_L^{\omega n}, T^{\omega n}_-
\eta_L^{\omega n} \lket_{\C^2}\:d\omega \:+\: \sum_{n \leq n_0} \oint_{D_\delta}
\left| \bra \eta_L,\, (Q_{n}\: S_\infty\:H\,(H-\omega_0)\, \eta_L) \ket \right| \:d\omega \:.
\end{eqnarray*}
Using the asymptotic form of the energy scalar product and the Hamiltonian
near the event horizon, it is obvious that the first term stays bounded
as~$L \to \infty$.
The second term is bounded according to Lemma~\ref{lemmaneg}.
For the contour integrals we can use the formula~(\ref{explicit})
on the real interval~$[\omega_0, \omega_0+\delta)$. Since Theorem~\ref{thmh1} gives us control
of the asymptotics of fundamental solution~$\acute{\phi}$ uniformly as~$u \to
-\infty$, it is clear that the integral over~$[\omega_0, \omega_0+\delta)$ is
bounded uniformly in~$L$. For the contour in the complex plane, we cannot work
with~(\ref{explicit}), but we must instead consider the formula for
the operator product~$Q_n\, S_\infty$ give in~\cite[Proposition~5.4]{FKSY}
together with the estimate for
the Green's function given in Lemma~\ref{lemmawrones} below.
\QED

\begin{Lemma} \label{lemmawrones}
For every~$\tilde{\omega} \in D_\delta$ with~$\omega \neq \omega_0$, there are
constants~$C, \epsilon>0$ and~$u_0 \in \R$ such the Green's function satisfies
for all~$\omega \in D_\delta \cap B_\epsilon(\tilde{\omega})$ the inequality
\[  |g(u,v)| \;\leq\; C \spc {\mbox{for all~$u,v \leq u_0$}}\:. \]
\end{Lemma}
{\Proof} It suffices to consider the case~${\mbox{Im}}\, \omega \leq 0$, because
the Green's function in the upper half plane is obtained simply by complex
conjugation. By symmetry, we can furthermore assume that~$u \leq v$.
Thus, according to~(\ref{Green}), we must prove the inequality
\[ \left| \frac{\acute{\phi}(u)\, \grave{\phi}(v)}{w(\acute{\phi}, \grave{\phi})}
\right| \;\leq\; C\spc {\mbox{for all~$u \leq v \leq u_0$}}\:. \]
According to Whiting's mode stability~\cite{W}, the Wronskian~$w(\acute{\phi},
\grave{\phi})$ has no zeros away from the real line, and thus by choosing~$\delta$
so small that~$B_\delta(\tilde{\omega})$ lies entirely in the lower half plane,
we can arrange that~$|w(\acute{\phi}, \grave{\phi})|$ is bounded
away from zero on~$B_\delta(\tilde{\omega})$. Hence our task is to bound the
factor~$|\acute{\phi}(u)\: \grave{\phi}(v)|$. Solving the defining equation for~$w(\acute{\phi}, \grave{\phi})$ for~$\grave{\phi}$ and integrating, we obtain
\[ \frac{\grave{\phi}}{\acute{\phi}} \Big|_{v}^{u_0} \;=\;
-w(\acute{\phi}, \grave{\phi}) \int_{v}^{u_0} \frac{du}{\acute{\phi}(u)^2} \:. \]
Substituting the identity
\[ \frac{1}{\acute{\phi}(u)^2} \;=\; \frac{e^{-2i \Omega u}}{2 i \Omega}\:
\frac{d}{du} \left( \frac{e^{2i \Omega u}}{\acute{\phi}(u)^2} \right) \:-\:
\frac{1}{2 i \Omega} \:\frac{d}{du} \left( \frac{1}{\acute{\phi}(u)^2} \right) , \]
the integral over the last term gives a boundary term,
\[ \int_{v}^{u_0} \frac{1}{2 i \Omega} \:\frac{d}{du} \left( \frac{1}{\acute{\phi}(u)^2} \right)
\;=\; \frac{1}{2 i \Omega}\left. \frac{1}{\acute{\phi}(u)^2}
\right|_{v}^{u_0}\:. \]
The integral over the other term can be estimated by
\[ \frac{1}{2 \Omega} \int_{v}^{u_0}
\left|e^{-2i \Omega u} \left( \frac{e^{2i \Omega u}}{\acute{\phi}(u)^2} \right)' \right| dv
\;\leq\; \frac{e^{-2 {\mbox{\scriptsize{Im}}}\, \Omega\: v}}{\Omega} \int_{v}^{u_0}
\left| \frac{(e^{-i \Omega u}\, \acute{\phi}(u))'}
{(e^{-i \Omega u}\, \acute{\phi}(u))^3} \right| dv \:. \]
Using the asymptotics~(\ref{abc1}) one sees that the last integrand vanishes
at the event horizon. From the series expansion for~$\acute{\phi}$, (\ref{pe}, \ref{iter}),
we see that this integrand decays even exponentially fast. Therefore, the
last integral is finite, uniformly in~$v$ and locally uniformly in~$\omega$.
Collecting all the obtained terms and using the known asymptotics~(\ref{abc1})
of~$\acute{\phi}$, the result follows.
\QED

\begin{Lemma} \label{lemmall}
For any~$\varepsilon>0$ we can, possibly by increasing~$n_0$,
arrange that for all~$L \geq 0$,
\[ \sum_{n > n_0} \int_{\omega_0}^0
\left| \lbra \eta^{\omega n}_L, T^{\omega n}_+ \zeta^{\omega n} \lket_{\C^2}
\right|\:d\omega \;\leq\; \varepsilon \:. \]
\end{Lemma}
{\Proof}
Again using positivity, it suffices to bound the terms
\[ \sum_{n > n_0} \int_{\omega_0}^0
\lbra \eta^{\omega n}_L, T^{\omega n}_+ \eta^{\omega n}_L \lket_{\C^2}\:d\omega
\quad {\mbox{and}} \quad
\sum_{n > n_0} \int_{\omega_0}^0
\lbra \zeta^{\omega n}, T^{\omega n}_+ \zeta^{\omega n} \lket_{\C^2} \:d\omega\: . \]
They can be treated similarly,
consider for example the first term. For any~$n_1 > n_0$,
\begin{eqnarray*}
\lefteqn{
\inf_{(\omega_0, 0)} \lambda_{n_1}^2 \sum_{n \geq n_1} \int_{\omega_0}^0
\lbra \eta^{\omega n}_L, T^{\omega n}_+ \eta^{\omega n}_L \lket_{\C^2}\:d\omega
\;\leq\; \sum_{n > n_0} \int_{\omega_0}^0
\lbra ({\mathcal{A}} \eta_L)^{\omega n}, T^{\omega n}_+
({\mathcal{A}} \eta_L)^{\omega n} \lket_{\C^2}\:d\omega } \\
&\leq& \bra {\mathcal{A}} \eta_L ,\, H\,(H-\omega_0)\,{\mathcal{A}} \eta_L \ket
+ \sum_{n > n_0} \int_{\omega_0}^0 \lbra ({\mathcal{A}} \eta_L)^{\omega n},
T^{\omega n}_- ({\mathcal{A}} \eta_L)^{\omega n} \lket_{\C^2}\:d\omega \spc\qquad \\
&& + \sum_{n \leq n_0} \oint_{D_\delta}
\left| \bra{\mathcal{A}} \eta_L,\, (Q_{n}\: S_\infty\:H\,(H-\omega_0)\,  {\mathcal{A}}
\eta_L)\ket \right| \:d\omega \:.
\end{eqnarray*}
Here~${\mathcal{A}}$ is the angular operator. When it acts on a test function,
we always get rid of the time-derivatives with the replacement
$i \partial_t \to H$.
We now argue as in the proof of Lemma~\ref{lemmale} (with~$\eta_L$ replaced
by~${\mathcal{A}} \eta_L$) and choose~$n_1$ sufficiently large.
\QED

\section{An Integral Representation on the Real Axis} \label{secsuper}
\setcounter{equation}{0}
We now use a causality argument together with the estimates of the previous
section to show that the radiant modes in Proposition~\ref{prpsrm}
must be absent. This will be a contradiction to the assumption that there
are radiant modes, ruling out the possibility that there are
radiant modes at all. This will lead us to an integral representation
of the propagator on the real axis.

Let us return to the setting of Proposition~\ref{prpsrm}. Choosing the
$\vartheta$-dependence of~$\eta$ such that it is orthogonal to the
angular wave functions~$(\Psi^{\omega_1 n}_a)_{n \leq n_0}$ except for one~$n$,
and choosing the~$u$-dependence of~$\eta$ such that it is orthogonal only
to one of the plane waves~$e^{\pm i (\omega_1-\omega_0) u}$, we can clearly arrange that
\beq \label{sigman}
\limsup_{L \to \infty} |\sigma(L)| \;=:\; \kappa \:>\: 0 \spc
{\mbox{where}} \spc \sigma(L) \;:=\; \sum_{n \leq n_0} \bra \eta^{\omega_1 n}_L,
\sigma^n \ket_{\C^2} \:.
\eeq
Furthermore, we choose~$\eta_L$ such that its support lies to the left of~${\mbox{supp}}\,
\Phi_0$, i.e.
\[ {\mbox{dist}}( {\mbox{supp}}\, \eta_L,\, {\mbox{supp}}\, \Phi_0) \;>\; L
\spc {\mbox{for all~$L \geq 0$}}. \]
Due to the finite propagation speed (which in the $(t,u)$-coordinates is equal to one),
\[ {\mbox{supp}}\, \eta_L \:\cap\: {\mbox{supp}}\, \Phi(t) \;=\; \emptyset
\spc {\mbox{if~$|t| \leq L$}}. \]
Hence for all~$L>0$,
\begin{eqnarray*}
0 &=& \frac{1}{2L} \int_{-L}^L e^{i \omega_1 t}\: \bra \eta_L,\: \Phi(t) \ket \\
&=& \frac{1}{2 \pi} \sum_{n \in \N}
\int_{-\infty}^\infty \frac{\sin((\omega-\omega_1)L)}{(\omega-\omega_1)L} \:
\lbra \eta_L^{\omega n}, T^{\omega n} \Phi_0^{\omega n} \lket_{\C^2}\:d\omega
\;+\; \sigma(L)\:.
\end{eqnarray*}
We apply Lemma~\ref{lemmaneg} and Lemma~\ref{lemmall} with
$\varepsilon=\kappa/(8 \pi)$ to obtain
\[ \frac{1}{2 \pi} \sum_{n > n_0}
\int_{\omega_0}^0 \left| \lbra \eta_L^{\omega n}, T^{\omega n}
\Phi_0^{\omega n} \lket_{\C^2} \right|\,d\omega \;\leq\; \frac{\kappa}{2}\:. \]
Furthermore, Lemma~\ref{lemmale} gives rise to the estimate
\[ \left| \int_{\R \setminus [\omega_0,0]}
\frac{\sin((\omega-\omega_1)L)}{(\omega-\omega_1)L} \:
\lbra \eta_L^{\omega n}, T^{\omega n} \Phi_0^{\omega n} \lket_{\C^2}\:d\omega \right|
\;\leq\; C\: \sup_{\R \setminus [\omega_0,0]}
\left|\frac{\sin((\omega-\omega_1)L)}{(\omega-\omega_1)L}\right| \:, \]
and since this supremum tends to zero as~$L \to \infty$, we conclude
that the expression on the left vanishes in the limit~$L \to \infty$.
Combining these estimates with~(\ref{sigman}), we obtain
\beq \label{contra}
\limsup_{L \to \infty}
\left| \sum_{n \leq n_0}
\int_{\omega_0}^0 \frac{\sin((\omega-\omega_1)L)}{(\omega-\omega_1)L} \:
\lbra \eta_L^{\omega n}, T^{\omega n} \Phi_0^{\omega n} \lket_{\C^2} \:d\omega \right|
\;\geq\; \pi \kappa\:.
\eeq
Since the matrices~$T^{\omega n}$ are continuous in~$\omega$ and the
fundamental solutions~$\Psi^{\omega n}_a(u)$ are according to Theorem~\ref{thmh1}
uniformly bounded as~$u \to -\infty$, there is a constant~$C$ such that
\[ \left| \lbra \eta_L^{\omega n}, T^{\omega n} \Phi_0^{\omega n} \lket_{\C^2} \right|
\;\leq\; C \spc {\mbox{for all~$L \geq 0$ and~$\omega \in (\omega_0,0),\; n \leq n_0$}}\:. \]
Hence we can apply Lebesgue's dominated convergence theorem on the left
of~(\ref{contra}) and take the limit~$L \to \infty$ inside the integral, giving zero.
This is a contradiction.

Since radiant modes have been ruled out, we know that the
Wronskian~$w(\acute{\phi}, \grave{\phi})$ has no zeros on the real
axis. Thus we can move all contours up to the real axis. This gives
the following integral representation for the propagator.

\begin{Thm} \label{thmrealrep}
For any initial data~$\Psi_0 \in C^\infty_0(\R \times S^2)^2$, the
solution of the Cauchy problem has the integral representation
\[ \Psi(t,r,\vartheta, \varphi)
\;=\; \frac{1}{2 \pi} \sum_{k \in \sZ} e^{-i k \varphi} \sum_{n \in
\sN} \int_{-\infty}^\infty \frac{d\omega}{\omega \Omega}\:e^{-i
\omega t} \sum_{a,b=1}^2 t^{k \omega n}_{ab}\: \Psi^a_{k \omega
n}(r,\vartheta)\; \bra \Psi^b_{k \omega n}, \Psi_0 \ket \] with the
coefficients~$t_{ab}$ as given by~(\ref{tabdef}, \ref{trans}). Here
the sums and the integrals converge
in~$L^2_{\mbox{\scriptsize{loc}}}$.
\end{Thm}

\section{Proof of Decay}
\setcounter{equation}{0}
As was recently pointed out to us by Thierry Daud{\'e}, there is an error
on the last page of the printed version of this paper (which is identical
with the previous online version). In order to keep the best possible agreement
between the printed and online versions, in this section we simply
give the erratum which
corrects the printed version.
The error was that the inequality~(8.3) in the printed version of this paper
cannot be applied to the function~$\Psi(t)$
because it does not satisfy the correct boundary conditions. This invalidates
the last two inequalities of the printed version, and thus the proof of decay is incomplete.
We here fill the gap using a different
method. At the same time, we will clarify in which sense the sum over the angular
momentum modes converges in~\cite[Theorem~1.1]{FKSY} and Theorem~\ref{thmrealrep},
an issue which so far was not treated in sufficient detail.
The arguments so far certainly yield
{\em{weak}} convergence in~$L^2_{\mbox{\scriptsize{loc}}}$; here we will
prove strong convergence. 

Our method here is to split the wave function into the high and low energy components.
For the high energy component, we show that the $L^2$-norm of the wave function
can be bounded by the energy integral,
even though the energy density need not be everywhere positive (Section~\ref{sec1}).
For the low energy component we refine our ODE techniques (Section~\ref{sec2}).
Combining these arguments with a Sobolev estimate and the Riemann-Lebesgue lemma
completes the proof (Section~\ref{sec3}).

We begin by considering the integral representation of Theorem~\ref{thmrealrep},
for fixed~$k$ and a finite number~$n_0$ of angular momentum modes,
\beq \label{intrep2}
\Phi^{n_0}(t,r,\vartheta) \;:=\; \int_{-\infty}^\infty
d\omega\; e^{-i \omega t}\, \hat{\Phi}^{n_0}(\omega,r,\vartheta)\:,
\eeq
where for notational convenience we have omitted the $\varphi$-dependence
(i.e.\ the factor~$e^{-i k \varphi}$), and~$\hat{\Phi}^{n_0}(\omega)$ is
defined by
\[ \hat{\Phi}^{n_0}(\omega,r,\vartheta) \;=\; \frac{1}{2 \pi} \;\frac{1}{\omega \Omega} \sum_{n=1}^{n_0}
\sum_{a,b=1}^2 t^{\omega n}_{ab}\: \Phi^a_{\omega
n}(r,\vartheta)\; \bra \Psi^b_{\omega n}, \Psi_0 \ket \]
(as in~\cite{FKSY}, we always denote the scalar wave function by~$\Phi$,
whereas~$\Psi=(\Phi, \partial_t \Phi)$ is a two-component vector).
We recall that for large~$\omega$, the WKB-estimates of~\cite[Section~6]{FKSY}
ensure that the fundamental solutions~$\Phi^b_{k \omega n}$ go over to plane
waves, and thus, since the initial data~$\Psi_0$ is smooth and compactly supported,
the function~$\hat{\Phi}^{n_0}(\omega,r,\vartheta)$ decays rapidly in~$\omega$
(for details on this method see~\cite[proof of Theorem~6.5]{K}).
As a consequence, $\Phi^{n_0}$ and its derivatives are, for~$r$
and~$\vartheta$ in any compact set, uniformly bounded in time.
Our goal is to obtain estimates uniform in~$n_0$, justifying that,
as $n_0 \to \infty$, $\Phi^{n_0}$ converges in $L^2_{\mbox{\scriptsize{loc}}}$
to the solution of the wave equation.

To arrange the energy splitting we choose for a given parameter~$J>0$,
a positive smooth function~$\chi_\Hp$ supported on~$(J, \infty)$
with~$\chi_\Hp|_{[2J, \infty)} \equiv 1$. We define~$\chi_\Hm$
by~$\chi_\Hm(\omega) = \chi_\Hp(-\omega)$ and set~$\chi_\LE = 1-\chi_\Hp - \chi_\Hm$.
We introduce the
high-energy contributions~$\hat{\Phi}^{n_0}_\Hpm$ and the
low-energy contribution~$\hat{\Phi}^{n_0}_\LE$ by
\[ \hat{\Phi}^{n_0}_\Hpm(\omega, r, \vartheta) \;=\; \chi_\Hpm(\omega)\,
\hat{\Phi}^{n_0}(\omega, r, \vartheta) \:,\qquad
\hat{\Phi}^{n_0}_\LE(\omega, r, \vartheta) \;=\; \chi_\LE(\omega)\,
\hat{\Phi}^{n_0}(\omega, r, \vartheta)\:. \]

\subsection{$L^2$-Estimates of the High-Energy Contribution} \label{sec1}
We recall from~\cite[(2.5)]{FKSY} that the energy density
of a wave function~$\Phi$ in the Kerr geometry is given by
\begin{eqnarray}
{\cal{E}}(\Phi) &=&
\left({\frac{(r^{2}+a^{2})^{2}}{\Delta}} - a^{2}\,\sin^{2}\vartheta \right) \left|\partial_{t} \Phi \right|^{2}+\Delta
\left|\partial_{r}\Phi \right|^{2} \nonumber \\
&&+\sin^{2}\vartheta \left|\partial_{\cos
\vartheta}\Phi \right|^{2}+\left(
{\frac{1}{\sin^{2}\vartheta}}-{\frac{a^{2}}{\Delta}}\right) k^2\, \left|\Phi \right|^{2} .
\label{energydensity}
\end{eqnarray}
Note that the energy density need not be positive due to the last term.
However, the next theorem shows that the energy integral
\[ E(\Phi) \;:=\; \int_{r_1}^\infty dr \int_{-1}^1 d\cos \vartheta\;
{\cal{E}}(\Phi(t)) \]
(which is independent of time due to energy conservation),
in the high-energy region is both positive and can be bounded from
below by the $L^2$-norm. In what follows, we only consider~$\Phi^{n_0}_\Hp$
because~$\Phi^{n_0}_\Hm$ can be treated similarly.

\begin{Thm} \label{thm1}
There exists a positive constant~$J_0$
(depending only on~$k$, but independent of~$n_0$ and~$\Psi_0$),
such that for all~$J \geq J_0$ the following inequality holds for every~$t$:
\[ E(\Phi^{n_0}_\Hp) \;\geq\; \frac{J^2}{2} \int_{r_1}^\infty \frac{(r^2+a^2)^2}{\Delta}\:
dr \int_{-1}^1 d\cos \vartheta\;
|\Phi^{n_0}_\Hp(t)|^2\:. \]
\end{Thm}

The remainder of this section is devoted to the proof of Theorem~\ref{thm1}.
We begin with the following lemma.
\begin{Lemma} \label{lemma1}
Let $g, \Phi$ be measurable functions,
$g$ real and~$\Phi$ complex, such that
$\Phi$ and $g \Phi$ are in~$L^1(\R)$. Then
\[ \int_\R d\omega \int_\R d\omega'\;
\min \!\Big( g(\omega), g(\omega') \!\Big)\;
\Phi(\omega) \, \overline{\Phi(\omega')}
\;\geq\; \inf g\;
\left| \int_\R \Phi \right|^2\:. \]
\end{Lemma}
{\Proof} Using a standard approximation argument, it suffices to consider the case that~$g$ and~$\Phi$
are simple functions of the form
\[ g(\omega) \;=\; \sum_{a=1}^A g_a\: \chi(K_a)\:,
\qquad \Phi(\omega) \;=\; \sum_{a=1}^A \Phi_a\: \chi(K_a) \:, \]
where $\chi(K_a)$ is the characteristic function of the
set~$K_a$, and~$(K_a)_{a=1,\ldots,A}$ forms a partition of~$\R$. Then the
above inequality reduces to
\[ \sum_{a,b=1}^A \min(g_a, g_b)\; \Phi_a\,|K_a| \; \overline{\Phi_b}
\,|K_b| \;\geq\; \min g \sum_{a,b=1}^A \, \Phi_a\,|K_a| \; \overline{\Phi_b}
\,|K_b| \:. \]
In the case~$A=2$ and~$g_1 \leq g_2$, this inequality follows immediately from
the calculation
\[  g_1\,|c_1|^2 + g_1\: (c_1\, \overline{c_2} + \overline{c_1}\, c_2) + g_2\, |c_2|^2
\;\geq\;  g_1\,|c_1|^2 + g_1\: (c_1\, \overline{c_2} + \overline{c_1}\, c_2) + g_1\, |c_2|^2
\;=\; g_1\: |c_1 + c_2|^2\:, \]
where~$c_a := \Phi_a\, |K_a|$.
In the case~$A=3$ and~$g_1 \leq g_2 \leq g_3$, we get
\begin{eqnarray*}
\lefteqn{ g_1 \left( |c_1|^2 + 2 {\mbox{Re}} (c_1 \overline{c_2} + c_1 \overline{c_3} ) \right)
+ g_2 \left( |c_2|^2 +  2 {\mbox{Re}} (c_2 \overline{c_3}) \right) + g_3 \, |c_3|^2 } \\
&\geq& g_1 \left( |c_1|^2 + 2 {\mbox{Re}} (c_1 \overline{c_2} + c_1 \overline{c_3} ) \right)
+ g_2 \left| c_2 +  c_3 \right|^2 \;\geq\; g_1 \left|c_1 + c_2 + c_3 \right|^2\:.
\end{eqnarray*}
The general case is similar.
\QED

The next lemma bounds the~$L^2$-norm of~$\Phi^{n_0}_\Hp$ and its
partial derivatives by a constant depending on~$n_0$ and $\Psi_0$.
\begin{Lemma} \label{lemma2}
There is a constant~$C=C(n_0, \Psi_0)$ such that for every~$t$,
\[ \int_{r_1}^\infty \frac{(r^2+a^2)^2}{\Delta}\:
dr \int_{-1}^1 d\cos \vartheta\;
\Big( |\Phi^{n_0}_\Hp(t)|^2 + |\partial_r \Phi^{n_0}_\Hp(t)|^2
+ \sum_{k=1}^3|\partial^k_t \Phi^{n_0}_\Hp(t)|^2 \Big)
\;\leq\; C \:. \]
\end{Lemma}
{\Proof} It suffices to consider one angular momentum mode.
For notational simplicity we omit the angular dependence.
Since $\Phi^{n_0}_\Hp$ and its derivatives are
locally pointwise bounded uniformly in time, it follows that
their $L^2$-norms on any compact set are bounded in time.

Near the event horizon, we work with the fundamental solution~$\acute{\phi}$
in the Regge-Wheeler variable~$u$ (see~\cite[(2.18, 5.2)]{FKSY}
and Section~\ref{secholo}. Then for any sufficiently small~$u_0$,
the integral of~$|\Phi|^2$ over the region~$u<u_0$ can be written as
\[ \int_{-\infty}^{u_0} du \:|\phi|^2\:, \]
where it is now convenient to write our integral representation~(\ref{intrep2}) in the form
\[ \phi(t,u) \;=\; \int_{-\infty}^\infty d\omega \left(h_+(\omega)\: \acute{\phi}_\omega(u)
+ h_-(\omega)\: \overline{\acute{\phi}_\omega(u)} \right) e^{-i \omega t} \:. \]
Here the functions~$h_\pm$ have rapid decay and, as they are supported away from the
set~$\{0, \omega_0\}$, they are also smooth (see Section~\ref{secholo}).
Using the Jost representation~(\ref{pe}, \ref{iter}), the function~$\phi(t,u)$ can be decomposed as
\[ \phi(t,u) \;=\; \phi_+(t,u) + \phi_-(t,u) + \rho(t,u) \]
where
\begin{eqnarray*}
\phi_\pm(t,u) &:=& \int_{-\infty}^\infty d\omega\: h_\pm(\omega)\: e^{\pm i (\omega-\omega_0) u - i \omega t} \\
|\rho(t,u)| &\leq& C\, e^{\gamma u} \quad {\mbox{for all~$t$, where~$C, \gamma>0$}}.
\end{eqnarray*}
Note that the smoothness of~$h_\pm$ implies that~$\phi_\pm$
decay rapidly in~$u$.
From the exponential decay of the factor~$e^{\gamma u}$ it is obvious that
the~$L^2$-norm of~$\rho$ is bounded uniformly in~$t$. The~$L^2$-norms of~$\phi_\pm$
can be estimated as
\[ \int_{-\infty}^{u_0} |\phi_\pm(t,u)|^2 du \;\leq\; \int_{-\infty}^\infty |\phi_\pm(t,u)|^2 du
\;=\; \int_{-\infty}^\infty |\phi_\pm(0,u')|^2 du' \;=:\; c\:, \]
where~$u' = u \mp t$.

Near infinity, we work similarly with the fundamental solutions~$\grave{\phi}$~(\ref{abc2}).
Again using the Jost representation (see~(pe2) and Lemma~\ref{lemmah2}),
we get terms depending only on~$t\pm u$ as well as error terms which decay like~$1/u$
and are thus in~$L^2$.

The time derivatives can be treated in the same way, since a time derivative merely gives a factor of~$\omega$ which can be absorbed into~$h_\pm$.
For the spatial derivatives we use similarly the estimates for
the first derivatives of the Jost functions.
\QED

Our next step is to decompose the energy integral into a convenient form.
To do this, we introduce a positive mollifier $\alpha \in C^\infty_0([-1,1])$
with the properties~$\alpha(-\omega)=\alpha(\omega)$ and
$\int \alpha(\omega) d\omega =1$. We define the function~$\Gamma(\omega-\omega')$ by
mollifying the Heaviside function~$\Theta$,
\beq \label{mH}
\Gamma(\omega-\omega') \;=\; (\Theta * \alpha)(\omega-\omega') \:.
\eeq
We now substitute the Fourier representation of~$\Phi^{n_0}_\Hp$ into the formula
for the energy density~(\ref{energydensity}). For simplicity we omit the indices~$n_0$ and~$H+$
in what follows. Omitting the first positive summand in~(\ref{energydensity}),
we get the inequality
\begin{eqnarray}
\lefteqn{ {\cal{E}}(\Phi)(t,r,\vartheta) \;\geq\;
\int_{-\infty}^\infty d\omega
\int_{-\infty}^\infty d\omega' \; e^{-i (\omega-\omega')t} } \nonumber \\
&\times \bigg\{\!\!\!\! & \left(
{\frac{1}{\sin^{2}\vartheta}}-{\frac{a^{2}}{\Delta}}\right) k^2\:
\hat{\Phi}(\omega)\, \overline{\hat{\Phi}(\omega')} \label{LO} \\
&&+\Gamma(\omega-\omega') \left( \Delta\, \partial_{r}\hat{\Phi}(\omega) \, \overline{\partial_r \hat{\Phi}(\omega')}
+ \sin^{2}\vartheta \,\partial_{\cos \vartheta}\hat{\Phi}(\omega) \, \overline{\partial_{\cos \vartheta}
\hat{\Phi}(\omega')} \right) \label{IR} \\
&&+(1-\Gamma)(\omega-\omega') \left( \Delta\, \partial_{r}\hat{\Phi}(\omega) \, \overline{\partial_r \hat{\Phi}(\omega')} + \sin^{2}\vartheta \,\partial_{\cos \vartheta}\hat{\Phi}(\omega) \, \overline{\partial_{\cos \vartheta}
\hat{\Phi}(\omega')} \right) \bigg\}. \label{IL} 
\end{eqnarray}
We multiply by a positive test function~$\eta(u) \in C^\infty_0(\R)$ and
integrate over $r$ and~$\cos \vartheta$.
Integrating by parts in~(\ref{IR}) and~(\ref{IL}) to the right and left, respectively,
we can use the wave equation
\begin{eqnarray*}
\lefteqn{ \left[ -\frac{\partial}{\partial r}\Delta\frac{\partial}{\partial r}
-\frac{1}{\Delta} \left( (r^{2}+a^{2}) \omega+ak \right)^{2} \right. } \\
&& \left. -\frac{\partial}{\partial \cos \vartheta} \sin^{2}\vartheta
\frac{\partial}{\partial \cos \vartheta}
+\frac{1}{\sin^{2}\vartheta}(a\omega \sin^{2}\vartheta + k
)^{2} \right] \hat{\Phi}(\omega) \;=\; 0
\end{eqnarray*}
to obtain
\begin{eqnarray}
\lefteqn{ \hspace*{-1.5cm} \int_{r_1}^\infty dr \int_{-1}^1 d\cos \vartheta\; \eta(u)\,
\Big((\ref{IR}) + (\ref{IL}) \Big) \;=\; \int_{r_1}^\infty dr \int_{-1}^1 d\cos \vartheta
\int_{-\infty}^\infty d\omega
\int_{-\infty}^\infty d\omega' \; e^{-i (\omega-\omega')t} } \nonumber \\
&\times \bigg\{\!\!\!\! & -\eta'(u)\: (r^2+a^2) \left(\Gamma\: \hat{\Phi}(\omega)\,
\overline{\partial_r \hat{\Phi}(\omega')} + (1-\Gamma)\: \partial_r \hat{\Phi}(\omega)\,
\overline{\hat{\Phi}(\omega')} \right) \label{C1} \\
&&+ \eta(u)\, \Big( \Gamma\, g(\omega') + (1-\Gamma)\, g(\omega)
\Big)\: \hat{\Phi}(\omega)\, \overline{\hat{\Phi}(\omega')}
\bigg\} \: , \label{C21}
\end{eqnarray}
where
\[ g(\omega, r, \vartheta) \;=\; \frac{1}{\Delta} \left( (r^{2}+a^{2}) \omega+ak \right)^{2} -
\frac{1}{\sin^{2}\vartheta}(a\omega \sin^{2}\vartheta + k )^{2} \:, \]
and we used that
\[ \frac{d}{dr} \eta(u) \;=\; \eta'(u)\: \frac{r^2+a^2}{\Delta}\:. \]
We interchange the orders of integration of the spatial and frequency integrals and
let~$\eta$ tend to the constant function one. In the term corresponding to~(\ref{LO}),
Lemma~\ref{lemma2} allows us to pass to the limit. In~(\ref{C1}, \ref{C21}) the situation is a bit more involved due to the factors of~$\Gamma$.
However, since multiplication by~$\Gamma(\omega-\omega')$
corresponds to convolution with its Fourier transform~$\check{\Gamma}(t)$,
we can again apply Lemma~\ref{lemma2} and pass to the limit using Lebesgue's dominated convergence theorem. To make this method more precise,
let us show in detail that the expression
\beq \label{exexp}
\int_{r_1}^\infty dr \int_{-1}^1 d\cos \vartheta
\int_{-\infty}^\infty d\omega
\int_{-\infty}^\infty d\omega' \; e^{-i (\omega-\omega')t} \:
\Big( 1-\eta(u) \!\Big)\:\Gamma(\omega-\omega') \:g(\omega')\:
\hat{\Phi}(\omega)\: \overline{\hat{\phi}(\omega')}
\eeq
tends to zero as~$\eta$ converges to the constant function one.
Rewriting the factor~$\Gamma$ with a time convolution, we obtain
the expression
\[ \int_{-\infty}^\infty d\tau\; \check{\Gamma}(\tau)  \:F(\tau)\]
where~$F$ and~$\check{\Gamma}$ are defined by
\begin{eqnarray*}
F(\tau) &=& \int_{r_1}^\infty dr \int_{-1}^1 d\cos \vartheta\:(1-\eta(u))\:
\Phi(t-\tau,r,\vartheta)\: (\check{g}*\overline{\Phi})(t-\tau, r,
\vartheta) \\
\check{\Gamma}(\tau) &=& \frac{1}{2 \pi} \int_{-\infty}^\infty
\Gamma(b)\: e^{-i b \tau}\: db \:.
\end{eqnarray*}
Writing~$g$ as a polynomial in~$\omega$,
\beq  g(\omega) \;=\; g_0 + g_1\; \omega + g_2\, \omega^2 \:, \label{gpoly} \eeq
where
\beq g_0 \;=\; \frac{a^2 k^2}{\Delta} - \frac{k^2}{\sin^2 \vartheta} \:,\quad
g_1 \;=\; 2 a k \left( \frac{r^2+a^2}{\Delta} - 1 \right) \:,\quad
g_2 \;=\; \frac{(r^2+a^2)^2}{\Delta} - a^2 \sin^2 \vartheta \:, \label{g012}
\eeq
the function~$\check{g}*\overline{\Phi}$ can be expressed explicitly
in terms of~$\overline{\Phi}$ and its time derivatives of order at most two.
In order to compute~$\check{\Gamma}$, we first note that
the Fourier transform of the Heaviside function~$\Theta$ is
\beq \label{chT}
\check{\Theta}(\tau) \;=\; \frac{1}{2 \pi}
\left(-i \: \frac{\mbox{PP}}{\tau} + \pi \delta(\tau) \right) ,
\eeq
where~``PP'' denotes the principal part.
Using~(\ref{mH}) together with the fact that convolution in momentum space
corresponds to multiplication in position space, we find that
\beq \label{chG}
\check{\Gamma}(\tau) \;=\;
\left(-i \: \frac{\mbox{PP}}{\tau} + \pi \delta(\tau) \right)
\check{\alpha}(\tau)\:,
\eeq
where~$\check{\alpha}$ is a Schwartz function
with~$\check{\alpha}(0)=(2 \pi)^{-1}$.
According to Lemma~\ref{lemma2}, the function~$F$ is uniformly bounded,
\[ |F(\tau)| \;\leq\; \sup|\eta| \quad {\mbox{for all~$\tau \in \R$}}. \]
Using the rapid decay of~$\check{\Gamma}$, we can for any given~$\varepsilon$
choose a parameter~$L>0$ such that
\[ \left| \int_{\R \setminus [-L,L]} \check{\Gamma}(\tau)\, F(\tau) \right|
\;\leq\; \varepsilon\:\sup|\eta|\:. \]
On the interval~$[-L,L]$, on the other hand, the singularity
of~$\hat{\Gamma}$ at~$\tau=0$ can be controlled by at most first
derivatives of~$F$, and thus for a suitable constant~$C=C(L)$,
\[ \left| \int_{-L}^L \check{\Gamma}(\tau)\, F(\tau) \right| \;\leq\;
2 C\: \sup_{[-L,L]} (|F| + |F'|)\:. \]
Using the rapid decay of~$\Phi$ and its time derivatives
in~$u$, locally uniformly in~$\tau$,
we can make~$\sup_{[-L,L]}(|F|+|F'|)$ as small as we like.
This shows that~(\ref{exexp}) really tends to zero as~$\eta$ 
goes to the constant function one.

We conclude that
\begin{eqnarray}
E(\Phi^{n_0}_\Hp) &\geq& \int_{r_1}^\infty dr \int_{-1}^1 d\cos \vartheta
\int_{-\infty}^\infty d\omega
\int_{-\infty}^\infty d\omega' \; e^{-i (\omega-\omega')t} \nonumber \\
&\times \bigg\{\!\!\!\! & 
\left( \min (g(\omega), g(\omega')) + {\frac{k^2}{\sin^{2}\vartheta}}-{\frac{a^{2}k^2}{\Delta}}
\right) \hat{\Phi}(\omega)\, \overline{\hat{\Phi}(\omega')} \label{M} \\
&&+ \Big( \Gamma\, g(\omega') + (1-\Gamma)\, g(\omega) - \min(g(\omega), g(\omega')) \Big)
\hat{\Phi}(\omega)\, \overline{\hat{\Phi}(\omega')} \; \bigg\} \: . \label{C2}
\end{eqnarray}
We apply Lemma~\ref{lemma1} to obtain
\[ (\ref{M}) \;\geq\;
\int_{r_1}^\infty dr \int_{-1}^1 d\cos \vartheta\,
\left(\inf_{\omega \geq J} g(\omega) 
+ {\frac{k^2}{\sin^{2}\vartheta}}-{\frac{a^{2}k^2}{\Delta}} \right)
\:|\Phi(t,r,\vartheta)|^2 \:. \]
Using the explicit form of~$g$, (\ref{gpoly}),
we find that for sufficiently large~$J$,
\beq (\ref{M}) \;\geq\; 
\frac{J^2}{2} \int_{r_1}^\infty \frac{(r^2+a^2)^2}{\Delta}\: dr \int_{-1}^1 d\cos \vartheta\;
|\Phi(t,r,\vartheta)|^2 \:. \label{Mgood}
\eeq

It remains to control the term~(\ref{C2}).
We write the $\omega, \omega'$-integral of~(\ref{C2})
in the form
\[ B \;:=\; \int_{-\infty}^\infty d\omega \int_{-\infty}^\infty d\omega' \; e^{-i (\omega-\omega')t}
h(\omega, \omega')\: \hat{\Phi}(\omega)\, \overline{\Phi(\omega')} \:, \]
where
\[ h(\omega, \omega') \;=\; \Gamma\, g(\omega') + (1-\Gamma)\, g(\omega) - \min(g(\omega), g(\omega'))\:. \]
Introducing the variables $a=\frac{1}{2}(\omega+\omega')$ and~$b=\frac{1}{2}(\omega-\omega')$,
and using that~$g(\omega)$ is a polynomial in~$\omega$, a short calculation
yields
\[ h(a+b, a-b) \;=\; \left(g_1 + 2 g_2\, a \right)\: S(2b) \qquad {\mbox{where}} \qquad
S(b) \;:=\; b\,\Big(\Theta(b) - \Gamma(b) \Big)\:. \]
Using~(\ref{chT}, \ref{chG}) together with the fact that the factor~$b$
corresponds to a derivative in position space, we obtain
\[ \check{S}(\tau) \;=\; \frac{1}{2\pi} \frac{d}{d \tau} \left( \frac{1-2 \pi \check{\alpha}(\tau)}{\tau} \right)
\;=\; -\frac{d}{d \tau} \int_0^1 \alpha'(s \tau)\:ds \:. \]
This is a smooth function which decays quadratically at infinity; in particular, it is integrable.

We thus obtain for the Fourier transform of~$h$ the explicit formula
\begin{eqnarray*}
\check{h}(\tau, \tau') &=& \frac{1}{(4 \pi)^2} \int_{-\infty}^\infty d\omega
\int_{-\infty}^\infty d\omega'\: h(\omega, \omega')\: e^{-i(\omega t - \omega' t')} \\
&=& \Big( g_1 \delta(\tau-\tau') + 2 i g_2\: \delta'(\tau-\tau') \Big)
\check{S}\!\left(\frac{\tau+\tau'}{2} \right)\:.
\end{eqnarray*}
Using Plancherel for distributions, we obtain
\begin{eqnarray*}
B &=& \int_{-\infty}^\infty d\tau \int_{-\infty}^\infty d\tau'\:
\check{h}(\tau, \tau')\: \Phi(t-\tau)\, \overline{\Phi(t-\tau')} \\
&=& \int_{-\infty}^\infty d\tau\, \check{S}(\tau) \Big[
g_1\: \Phi(t-\tau)\, \overline{\Phi(t-\tau)} \\
&& \spc\qquad\;\; + i g_2 \left(
\partial_t \Phi(t-\tau)\, \overline{\Phi(t+\tau)} - \Phi(t-\tau)\, \overline{\partial_t \Phi(t+\tau)}
\right) \Big] \:.
\end{eqnarray*}
Integrating over space, we can use the explicit formulas for~$g_1$ and~$g_2$
and apply Lemma~\ref{lemma2} to obtain
\[ (\ref{C2}) \;=\; \int_{r_1}^\infty
dr \int_{-1}^1 d\cos \vartheta\; B \;\geq\; -C(n_0, \Psi_0)
\int_{-\infty}^\infty|\check{S}(\tau)| \: d\tau \:. \]
We now let~$\alpha$ tend to the Dirac delta; then~$\check{\alpha}$ tends to the
constant function~$(2 \pi)^{-1}$. As a consequence, the $L^1$-norm of~$\check{S}$
tends to zero, and thus~(\ref{C2}) becomes positive in this limit.
Hence the energy is bounded from below by~(\ref{Mgood}). This concludes the
proof of Theorem~\ref{thm1}.

\subsection{Pointwise Estimates for the Low-Energy Contribution} \label{sec2}
The low-energy contribution can be written as
\[ \Phi^{n_0}_\LE (t,r,\vartheta) \;=\; \frac{1}{2 \pi}  \sum_{n=1}^{n_0} \int_{-\infty}^\infty
\frac{d\omega}{\omega \Omega}\; e^{-i \omega t}\; \chi_\LE(\omega)
\sum_{a,b=1}^2 t^{\omega n}_{ab}\: \Phi^a_{\omega
n}(r,\vartheta)\; \bra \Psi^b_{\omega n}, \Psi_0 \ket \:. \]
We now derive pointwise estimates for the large angular momentum  modes.
\begin{Thm} \label{thm2}
For any~$u_0 < u_1$ there is a constant~$C>0$ such that for
all~$\omega \in (-2J, 2J) \setminus \{\omega_0, 0\}$ and for
all~$u, u' \in (u_0, u_1)$,
\beq \label{nsum}
\sum_{n=1}^\infty \left| \frac{1}{\Omega} 
\sum_{a,b=1}^2 t_{ab}\: \phi^a(u)
\: \phi^b(u') \right| \;<\; C\:.
\eeq
\end{Thm}
{\Proof} From~(\ref{tabdef}) the coefficients~$t_{ab}$ have the explicit form
\[ T \;:=\; (t_{ab}) \;=\; \left(\! \begin{array}{cc} \displaystyle 1 + {\mbox{Re}}\, \frac{\alpha}{\beta}
& \displaystyle -{\mbox{Im}}\, \frac{\alpha}{\beta} \\[1em]
\displaystyle -{\mbox{Im}}\, \frac{\alpha}{\beta} & \displaystyle 1 - {\mbox{Re}}\, \frac{\alpha}{\beta}
\end{array}\! \right) , \]
where the transmission coefficients~$\alpha$ and~$\beta$ are defined by
\[ \grave{\phi} \;=\; \alpha\: \acute{\phi} + \beta\: \overline{\acute{\phi}} \:, \]
and~$\phi_1 = {\mbox{\rm{Re}}}\, \acute{\phi}$, $\phi_2 = {\mbox{\rm{Im}}}\, \acute{\phi}$.
The estimates of Section~\ref{sec4.3} are obviously valid for~$\omega$ in any bounded
set; in particular for~$\omega \in (-2J, 2J) \setminus \{\omega_0, 0\}$. We use these estimates
in what follows, also using the same notation. We choose~$n_1$ so large that~$u_+ < u_0$,
and thus on the whole interval~$(u_0, 2u_1)$ the invariant disk estimates of Lemmas~\ref{lemmainv2}
and~\ref{lemma4.8} hold.

Rewriting the expression~$t_{ab} \phi^a \phi^b$ with the Green's function (see the proof of Lemma~\ref{lemma5.1}),
this expression is clearly invariant under the phase
transformation~$\acute{\phi} \to e^{i \vartheta} \acute{\phi}$. Thus we can arrange
that~$\acute{\phi}(2u_1)$ is real. Then the transmission coefficients are computed at~$u=2u_1$ by
\[ \left(\!\! \begin{array}{c} \grave{\phi} \\ \grave{\phi}' \end{array} \!\!\right)
\;=\; \left(\! \begin{array}{cc} \acute{\phi} & \overline{\acute{\phi}} \\
\acute{\phi}' & \overline{\acute{\phi}'} \end{array} \right)
\left(\!\! \begin{array}{c} \alpha \\ \beta \end{array} \!\!\right)
\;=\; \acute{\phi} \left(\! \begin{array}{cc} 1 & 1 \\
\acute{y} & \overline{\acute{y}} \end{array} \right)
\left(\!\! \begin{array}{c} \alpha \\ \beta \end{array} \!\!\right) . \]
We thus obtain
\[ \alpha \;=\; -\frac{\grave{\phi}}{2\,\acute{\phi}\, {\mbox{Im}}\, \acute{y}}
\left( \overline{\acute{y}} - \grave{y} \right) \Big|_{u=2 u_1}\,\qquad
\beta \;=\; \frac{\grave{\phi}}{2\,\acute{\phi}\,{\mbox{Im}}\, \acute{y}}
\left( \acute{y} - \grave{y} \right) \Big|_{u=2 u_1} \:. \]
Hence
\[ \left| 1 + \frac{\alpha}{\beta} \right| \;=\; 2\:
\frac{|{\mbox{Im}}\, \acute{y}|}{|\acute{y} - \grave{y}|} 
\;\leq\; \frac{4\, |\Omega|}{\acute{\rho}^2\: {\mbox{Re}}\, (\acute{y} - \grave{y})} \:, \]
where all functions are evaluated at~$u=2u_1$, and where we used the relation
\beq \label{ryrel}
\acute{\rho}^2 \;=\; \frac{|\Omega|}{{\mbox{Im}}\, \acute{y}} \:,
\eeq
which is an immediate consequence of~(\ref{5C})
and~$w(\acute{\phi}, \overline{\acute{\phi}})=2i \Omega$.
From Lemma~\ref{lemmaeasy} and~(\ref{corin1}) we know that
for~$\lambda$ sufficiently large,
\[ {\mbox{Re}}\, (\acute{y} - \grave{y}) \;\geq\; \frac{1}{C} \:. \]
Using the above formulas for~$t_{ab}$, we conclude that
\[ \left| \sum_{(a,b) \neq (2,2)} \frac{1}{\Omega}\: t_{ab}\: \phi^a(u)\, \phi^b(u') \right|
\;\leq\; 12\,C \:
\frac{\acute{\rho}(u)}{\acute{\rho}(2u_1)}\: \frac{\acute{\rho}(u')}{\acute{\rho}(2u_1)} . \]
The argument after~(\ref{inter2}) shows that the two factors on the right decay
like~$\exp(-\sqrt{\lambda}/c)$.

It remains to consider the case~$a=b=2$. Taking the imaginary part of the identity
\[ \acute{\phi}(u) \;=\; \acute{\phi}(2u_1)\: \exp \left( \int_{2u_1}^u \acute{y} \right) \]
and using that~$\acute{\phi}(2u_1)$ is real, we find that
\[ \acute{\rho}(u) \;=\; \acute{\rho}(2u_1)\:
\exp \left( \int_{2u_1}^u {\mbox{Re}}\,\acute{y} \right) \]
and thus
\[ |\phi_2(u)| \;=\; \acute{\rho}(u)\;  \left| \,\sin \left( \int_{2u_1}^u {\mbox{Im}}\, \acute{y} \right) \right|
\;\leq\; \acute{\rho}(u) \left| \int_u^{2u_1} {\mbox{Im}}\, \acute{y} \right| \:. \]
From~(\ref{corin0}) we see that~${\mbox{Re}}\, y$ is positive on~$[u,2u_1]$.
Using the relation~$\rho'/\rho={\mbox{Re}}\, y$, we conclude that~$\acute{\rho}$ is monotone increasing,
and~(\ref{ryrel}) yields that~${\mbox{Im}}\, \acute{y}$ is decreasing.
Hence, again using~(\ref{ryrel}),
\[ |\phi_2(u)| \;\leq\; \acute{\rho}(u)\: {\mbox{Im}}\, \acute{y}(u)\, (2u_1-u_0)
\;\leq\; \sqrt{|\Omega|\, {\mbox{Im}}\, \acute{y}(u_0) } \, (2u_1-u_0) \:. \]
Using the above estimates for~$\alpha/\beta$, we conclude that
\beq \label{t22term}
\left| \frac{1}{\Omega}\: t_{22}\: \phi^2(u)\, \phi^2(u') \right|
\;\leq\; 3  \,{\mbox{Im}}\, \acute{y}(u_0) \, (2u_1-u_0)^2 \:.
\eeq
The invariant region estimate of Lemmas~\ref{lemmainv2} and~\ref{lemma4.8} yield that
\[ {\mbox{Im}}\, \acute{y}(u_0) \;\leq\; c |\Omega|\: \exp \left( - \frac{7}{8} \int_{u_+}^{u_0} 
\sqrt{V} \right) \:, \]
and we conclude that~(\ref{t22term}) again decays like~$\exp(-\sqrt{\lambda}/c)$.

Since the eigenvalues~$\lambda_n$ scale quadratically in~$n$ (\ref{quadra}), the summands in~(\ref{nsum}) decay exponentially in~$n$.
\QED

This theorem gives us pointwise control of the low-energy contribution,
uniformly in time and in~$n_0$, locally uniformly in space.
To see this, we estimate the integral representation~(\ref{intrep2}) by
\[ |\Phi^{n_0}_\LE(t,r,\vartheta)| \;\leq\;
\frac{1}{2 \pi} \sum_{n=1}^{n_0}
\int_{-\infty}^\infty d\omega\: |\chi_L(\omega)|\:
\sum_{a,b=1}^2
\left| \frac{1}{\omega \Omega}\:
t_{ab}^{\omega n} \Phi^a_{\omega n}(r,\vartheta)
\: \bra \Psi^b_{\omega n}, \Psi_0 \ket \right| . \]
In order to control the factor~$\omega^{-1}$, we write
the energy scalar product on the right
in the form~\cite[(2.15)]{FKSY}, which involves an
overall factor~$\omega$. Now we can apply Theorem~\ref{thm2}.

\subsection{Decay in~$L^\infty_{\mbox{\scriptsize{loc}}}$} \label{sec3}
In this section we complete the proof of Theorem~\ref{thmmain}.
Let~$K \subset (r_1, \infty) \times S^2$ be a compact set.
The~$L^2$-norm of~$\Phi^{n_0}$ can be
estimated by
\[ \|\Phi^{n_0}(t)\|_{L^2(K)} \;\leq\; \|\Phi^{n_0}_\Hp(t)\|_{L^2(K)} + \|\Phi^{n_0}_\Hm(t)\|_{L^2(K)}
+ \|\Phi^{n_0}_\LE(t)\|_{L^2(K)} \:. \]
According to Theorems~\ref{thm1} and~\ref{thm2}, these norms are bounded uniformly in~$n_0$ and~$t$.
Furthermore, our estimates imply that the sequence~$\Phi^{n_0}(t)$ forms a Cauchy sequence in~$L^2(K)$.
To see this, we note that for any~$n_1, n_2$,
\[ \|\Phi^{n_1} - \Phi^{n_2}\|_{L^2(K)} \;\leq\; E \Big(\Phi^{n_1}_\Hp - \Phi^{n_2}_\Hp \!\Big)
 + E \Big( \Phi^{n_1}_\Hm - \Phi^{n_2}_\Hm \!\Big) + \|\Phi^{n_1}_\LE - \Phi^{n_2}_\LE\|_{L^2(K)} \:, \]
uniformly in~$t$.
The energy terms on the right are the sums of the energies of the individual angular momentum modes.
All the summands are positive due to Theorem~\ref{thm1}, and thus the energy terms become small
as~$n_1, n_2 \to \infty$.
The same is true for the last summand due to our OD estimates of Theorem~\ref{thm2}.
We conclude that~$\Phi^{n_0}(t)$ converges in~$L^2_{\mbox{\scriptsize{loc}}}$ as~$n_0 \to \infty$,
and the limit coincides
with the weak limit, which from~\cite{FKSY} we know to be the solution~$\Phi(t)$ of the
Cauchy problem.

To prove decay, given any~$\varepsilon>0$ we choose~$n_0$ such that~$\|\Phi(t) - \Phi^{n_0}(t)\|_{L^2(K)} <
\varepsilon$ for all~$t$. Since~$\hat{\Phi}^{n_0}$ is continuous in~$\omega$ and has
rapid decay, uniformly on~$K$, the Riemann-Lebesgue lemma yields that~$\Phi^{n_0}(t)$
decays in~$L^\infty(K) \subset L^2(K)$. Since~$\varepsilon$ is arbitrary, we conclude
that~$\Phi(t)$ decays in~$L^2(K)$.

Applying the same argument to the initial data~$H^n \Psi_0$, we conclude that
the partial derivatives of~$\Phi(t)$ also decay in~$L^2(K)$. The Sobolev
embedding~$H^{2,2}(K) \hookrightarrow L^\infty(K)$ completes the proof. \\

\noindent
{\em{Acknowledgments:}} 
We would like to thank Johann
Kronthaler and Thierry Daud{\'e} for helpful discussions.
We are grateful to the Vielberth
foundation, Regensburg, for support.


\begin{tabular}{lcl}
\\
Felix Finster & $\;\;\;\;$ & Niky Kamran\\
NWF I -- Mathematik && Department of Math.\ and Statistics \\
Universit{{\"a}}t Regensburg && McGill University \\
93040 Regensburg, Germany && Montr{\'e}al, Qu{\'e}bec \\
{\tt{Felix.Finster@mathematik}} && Canada H3A 2K6  \\
$\;\;\;\;\;\;\;\;\;\;\;\;\;\;$ {\tt{.uni-regensburg.de}}
&& {\tt{nkamran@math.McGill.CA}} \\
\\
Joel Smoller & $\;\;$ & Shing-Tung Yau \\
Mathematics Department && Mathematics Department \\
The University of Michigan && Harvard University \\
Ann Arbor, MI 48109, USA && Cambridge, MA 02138, USA \\
{\tt{smoller@umich.edu}} && {\tt{yau@math.harvard.edu}}
\end{tabular}


\begin{thebibliography}{99}
\bibitem{CY} V.\ Cardoso and S.\ Yoshida,``Superradiant
instabilities of rotating black branes and strings,''hep-th/0502206.
\bibitem{C} S.\ Chandrasekhar,``The mathematical theory of black
holes,'' {\em{Oxford University Press}} (1983)
\bibitem{AR} De Alfaro, Regge, ``Potential Scattering,''
{\em{North-Holland Publishing Company}}, Amsterdam (1965)
\bibitem{FKSY03} F.\ Finster, N.\ Kamran, J.\ Smoller, S.-T.\ Yau,
``The long-time dynamics of Dirac particles in the Kerr-Newman
black hole geometry,'' {\em{Adv.\ Theor.\ Math.\ Phys.}}\ {\bf 7} (2003) 25--52
\bibitem{FKSY} F.\ Finster, N.\ Kamran, J.\ Smoller, S.-T.\ Yau,
``An integral spectral representation of the propagator for the wave
equation in the Kerr geometry,'' gr-qc/0310024,  {\em{Commun.\ Math.\ Phys.}}
{\bf{260}} (2005) 257-298
\bibitem{FS} F.\ Finster, H.\ Schmid, ``Spectral estimates and non-selfadjoint
perturbations of spheroidal wave operators,'' math-ph/0405010,
{\em{J.\ Reine Angew.\ Math.}}\ {\bf{601}} (2006) 71-107 (2006)
\bibitem{KW}B.\ Kay, R.\ Wald, ``Linear stability of Schwarzschild under perturbations which are
nonvanishing on the bifurcation $2$-sphere,'' {\em{Classical Quantum
Gravity}}~{\bf{4}} (1987) 893--898.
\bibitem{K} J.\ Kronthaler, ``The Cauchy problem for the wave equation in the Schwarzschild geometry,'' gr-qc/0601131, {\em{J.\ Math.\ Phys.}} {\bf{47}} (2006) 042501
\bibitem{PT} W.H.\ Press, S.A.\ Teukolsky, ``Perturbations of a rotating black
hole. II. Dynamical stability of the Kerr metric,'' Astrophys.\ J.\ {\bf{185}}
(1973) 649
\bibitem{Price} R.H.\ Price, ``Nonspherical perturbations of relativistic
gravitational collapse I, scalar and gravitational perturbations,''
{\em{Phys.\ Rev.}}\ D (3) {\bf{5}} (1972) 2419--2438
\bibitem{W} B.\ Whiting, ``Mode stability of the Kerr black hole,''
{\em{J.\ Math.\ Phys.}}~{\bf{30}} (1989) 1301--1305
\bibitem{RW} T.\ Regge, J.A.\ Wheeler, ``Stability of the Schwarzschild
singularity,'' Phys.\ Rev.\ (2) {\bf{108}} (1957) 1063--1069
\end{thebibliography}
\end{document}